\documentclass[a4paper,11pt]{article}
\pdfoutput=1
\usepackage{jheppub}
\usepackage[T1]{fontenc}
\usepackage[utf8]{inputenc}

\usepackage{braket}
\usepackage{bbold}
\usepackage{subcaption}

\usepackage{amsthm}
\usepackage{amsmath}

\newtheorem{proposition}{Proposition}[section]

\graphicspath{ {./plots/} }

\renewcommand{\Im}{\operatorname{Im}}

\title{The Brownian loop soup stress-energy tensor}

\author[a,b,c]{Federico Camia}
\author[a]{Valentino F.\ Foit}
\author[a]{Alberto Gandolfi}
\author[d]{Matthew Kleban}

\affiliation[a]{Science Division, New York University Abu Dhabi, Saadiyat Island, Abu Dhabi, United Arab Emirates}
\affiliation[b]{Courant Institute of Mathematical Sciences, New York University, 251 Mercer Street, New York, NY 10012, USA}
\affiliation[c]{Department of Mathematics, Vrije Universiteit Amsterdam, De Boelelaan 1111, 1081 HV Amsterdam, The Netherlands}
\affiliation[d]{Center for Cosmology and Particle Physics, New York University, 726 Broadway, New York, NY 10003, USA}

\emailAdd{federico.camia@nyu.edu}
\emailAdd{foit@nyu.edu}
\emailAdd{albertogandolfi@nyu.edu}
\emailAdd{kleban@nyu.edu}

\abstract{The Brownian loop soup (BLS) is a conformally invariant statistical ensemble of random loops in two dimensions characterized by an intensity $\lambda>0$.
Recently, we constructed families of operators in the BLS and showed that they transform as conformal primary operators.
In this paper we provide an explicit expression for the BLS stress-energy tensor and compute its operator product expansion with other operators.  Our results are consistent with the conformal Ward identities and our previous result that the central charge is $c = 2 \lambda$.  In the case of domains with boundary we identify a boundary operator that has properties consistent with the boundary stress-energy tensor.
We  show that this operator generates local deformations of the boundary and that it is related to a boundary operator that induces a Brownian excursion starting or ending at its insertion point.
}
\begin{document}
\maketitle
\flushbottom

\section{Introduction} \label{sec:intro}

The Brownian loop soup (BLS) \cite{lawler2004brownian} is an ideal gas of Brownian loops in two dimensions with a distribution chosen so that it is invariant under local conformal transformations. The distribution depends on a single parameter, the intensity $\lambda > 0$.  The BLS is implicit in the work of Symanzik \cite{osti_4117149} on Euclidean quantum field theory, more precisely, in the representation of the partition function of Euclidean fields in terms of random paths that are locally statistically equivalent to Brownian motion. This representation can be made precise for the Gaussian free field, in which case the random paths are independent of each other and can be generated as a Poisson process. Viewed this way, the BLS consists of  $2 \lambda$ Gaussian free fields (a generally non-integer number).  

The BLS is closely related not only to Brownian motion and the Gaussian free field but also to the Schramm-Loewner Evolution (SLE) and Conformal Loop Ensembles (CLEs). It provides an interesting and useful link between Brownian motion, field theory, and statistical mechanics. Inspired and partly motivated by these connections, in \cite{Camia_2016,Camia_2020,camia2021scalar} we introduced and studied operators that compute properties of the BLS, discovering new families of conformal primary fields and analyzing the operator content of the emerging conformal field theory.

In this paper, we derive the bulk stress-energy tensor of the theory as well as the boundary stress-energy tensor in the upper half-plane. We also analyze the relation of the boundary stress-energy tensor to domain perturbations and boundary condition changing operators.

\subsection{Preliminary definitions} \label{sec:definitions}

The Brownian Loop Soup (BLS) is an ideal gas of planar loops. 
If $A$ is a set of loops, the partition function of the BLS restricted to loops from $A$ can be written as
\begin{align}
    {Z_A = \sum_{n=0}^{\infty} \frac{\lambda^n}{n!} \left( \mu^{\text{loop}}(A) \right)^n},
\end{align}
where $\lambda>0$ is a constant and $\mu^{\text{loop}}$ is a measure on planar loops called \emph{Brownian loop measure} and defined as
\begin{align} \label{brownian-loop-measure}
\mu^{\text{loop}} \equiv \int_{\mathbb C} \int_0^{\infty} \frac{1}{2 \pi t^2} \, \mu^{br}_{z,t} \, dt \, d{\bf A}(z),
\end{align}
where $\bf A$ denotes area and $\mu^{br}_{z,t}$ is the complex Brownian bridge measure with starting point
$z$ and duration $t$.\footnote{We note that the Brownian loop measure should be interpreted as a measure on ``unrooted''
loops, that is, loops without a specified starting point. Unrooted loops are equivalence classes of rooted
loops. The interested reader is referred to \cite{lawler2004brownian} for more details.}
$Z_A$ can be thought of as the grand canonical partition function of a system of loops with fugacity $\lambda$, and the BLS can be shown to be conformally invariant and to have central charge $c=2\lambda$, see \cite{lawler2004brownian,Camia_2016}.

As in \cite{camia2021scalar}, in this paper we will only be concerned with the of outer boundaries of Brownian loops. More precisely, given a planar loop $\gamma$ in $\mathbb C$, its outer boundary $\ell=\ell(\gamma)$ is the boundary of the unique infinite component of ${\mathbb C} \setminus \gamma$. 
Note that, for any planar loop $\gamma$, $\ell(\gamma)$ is always a simple closed curve, i.e., a closed loop without self-intersections. Hence, in this paper, we will work with collections ${\mathcal L}$ of simple loops $\ell$ which are the outer boundaries of the loops from a BLS and for us, with a slight abuse of terminology, a BLS will be a collection of simple loops.

Given a simple loop $\ell$, let $\bar\ell$ denote its \emph{interior}, i.e., the unique, bounded, simply connected component of $\mathbb{C} \setminus \ell$. In other words, a point $z$ belongs to $\bar\ell$ if $\ell$ disconnects $z$ from infinity, in which case we write $z \in \bar\ell$. In \cite{Camia_2016}, the authors studied the correlation functions of the \emph{layering operator}\footnote{In this paper we  use the terms \emph{field} and \emph{operator} interchangeably.}
\begin{align}
    V_{\beta}(z)=\exp\Big({i\beta\sum_{\ell: z \in \bar\ell} \sigma_{\ell}}(z)\Big),
\end{align}
where $\sigma_{\ell}$ are independent, symmetric, $(\pm 1)$-valued Boolean variables associated to the loops. One difficulty arises immediately due to the scale invariance of the BLS, which implies that the sum at the exponent is infinite with probability one. This difficulty can be overcome by imposing a short-distance cutoff $\delta>0$ 
on the diameter of loops\footnote{An additional infrared cutoff or a ``charge neutrality'' or ``charge conservation'' condition may be necessary in some circumstances --- we refer the interested reader to \cite{Camia_2016} for more details.} (essentially removing from the loop soup all loops with diameter smaller than $\delta$), which produces a cutoff version $V^{\delta}_{\beta}$ of $V_{\beta}$. The cutoff $\delta$ can be removed by rescaling $V^{\delta}_{\beta}$ by $\delta^{-2\Delta(\beta)}$, with $\Delta(\beta) = \frac{\lambda}{10}(1-\cos\beta)$, and sending $\delta \to 0$. When $\delta \to 0$, the $n$-point correlation functions of $\delta^{-2\Delta(\beta)}V^{\delta}_{\beta}$ converge to conformally covariant quantities \cite{Camia_2016}, showing that the limiting field is a scalar conformal primary field with real and positive conformal dimensions varying continuously as periodic functions of $\beta$, namely as
\begin{align}
    \Delta(\beta) = \bar \Delta(\beta) = \frac{\lambda}{10}(1-\cos\beta).
\end{align}
This limiting field is further studied in \cite{Camia_2020}, where its canonically normalized version is denoted by $\mathcal{O}_{\beta}$.\footnote{By \emph{canonically normalized} we mean that $\Braket{\mathcal{O}_{\beta}(z) \mathcal{O}_{-\beta}(z')}_{\mathbb C} = |z-z'|^{-4\Delta(\beta)}$.}

For a domain $D$ with a boundary $\partial D$ and a point $\zeta \in \partial D$, the \emph{boundary field} $\mathcal{B}^{\delta}_{\varepsilon}(\zeta)$ studied in this paper counts the number of loops $\ell$ with diameter at least $\delta$ passing within a short-distance $\varepsilon$ of $\zeta$. This is a boundary version of the \emph{edge field} studied in \cite{camia2021scalar} and discussed in Section \ref{sec:edge} below. As in \cite{camia2021scalar}, the cutoffs $\delta,\varepsilon>0$ are necessary to keep $\mathcal{B}^{\delta}_{\varepsilon}(\zeta)$ from being infinite or identically zero.
We show that they can be removed (i.e.\ sent to zero) by placing the field $\mathcal{B}^{\delta}_{\varepsilon}$ inside $n$-point correlation functions with $n \geq 2$ and renormalizing it by an appropriate power of $\varepsilon$.

\subsection{The edge counting operator} \label{sec:edge}

For the reader's convenience, in this section we provide a brief introduction to the \emph{edge counting operator} introduced in \cite{camia2021scalar}. Some of the properties of the edge counting operator will be essential to the analysis carried out in this paper.

The edge counting operator $\mathcal{E}(z)$, for $z$ in the \emph{interior} of a domain $D$, is  defined in \cite{camia2021scalar} as the limit
\begin{align}
\begin{split} \label{def:edge}
    {\mathcal E}(z) := \frac{\hat{c}}{\sqrt\lambda} \lim_{\varepsilon \to 0} \varepsilon^{-2/3} E_{\varepsilon}(z),
\end{split}
\end{align}
with
\begin{align}
\begin{split} \label{Edef}
    E_{\varepsilon}(z) := & \lim_{\delta \to 0} \left( N^{\delta}_{\varepsilon}(z) - \Braket{ N^{\delta}_{\varepsilon}(z) }_D \right) \\
    = & \lim_{\delta \to 0} \left( N^{\delta}_{\varepsilon}(z) - \lambda\mu^{\text{loop}}_D \left( \text{diam}(\ell)>\delta, \ell \cap B_{\varepsilon}(z) \neq \emptyset \right) \right).
\end{split}
\end{align}
Here $N^{\delta}_{\varepsilon}(z)$ counts the number of loops $\gamma$ with diameter\footnote{The diameter of a loop is the largest distance between any two point on the loop.} $\text{diam}(\gamma) = \text{diam}(\ell) $ larger than $\delta$ and  whose ``edge'' $\ell$ (the outer boundary) comes $\varepsilon-$close to $z$.  The angle brackets $\langle \cdot \rangle_D$ denote expectation with respect to the Brownian loop soup in $D$ (of fixed intensity $\lambda$),  $\mu^{\text{loop}}_D$ is the Brownian loop measure $\mu^{\text{loop}}$ restricted to $D$, i.e.\ the unique (up to a multiplicative constant) conformally invariant measure on simple planar loops \cite{2005math.....11605W}, and $B_{\varepsilon}(z)$ is the disk of radius $\varepsilon$ centered at $z$.

The subtraction of the mean in \eqref{Edef} is needed because the mean is divergent in the limit $\delta \to 0$, due to the scale invariance of the BLS. With this, the field $E_{\varepsilon}$ is well defined in the sense of correlation functions. More precisely, it is shown in \cite{camia2021scalar} that, for any collection of points $z_1,\ldots,z_n \in D$ at distance greater than $2\varepsilon$ from each other, with $n \geq 2$, the following limit exists:
\begin{align} \label{eq:E-correlations}
    \langle E_{\varepsilon}(z_1) \ldots E_{\varepsilon}(z_n) \rangle_D := \lim_{\delta \to 0} \langle E^{\delta}_{\varepsilon}(z_1) \ldots E^{\delta}_{\varepsilon}(z_n) \rangle_D .
\end{align}

The exponent $2/3$ used in \eqref{def:edge} is a consequence of the fact that the $\mu^{\text{loop}}_D$-measure of ``macroscopic'' loops coming to distance $\varepsilon$ of $z$ in the interior of $D$ tends to zero as $\varepsilon^{2/3}$ when $\varepsilon \to 0$. This can be understood using a deep connection \cite{2005math.....11605W} between the Brownian loop measure $\mu^{\text{loop}}_D$ and a conformally invariant model of non-simple loops, called CLE$_6$, which emerges from the scaling limit of critical two-dimensional percolation \cite{Camia_2006}. Using this connection, the exponent $2/3$ is directly related to the probability of a three-arm event in the bulk in two-dimensional percolation \cite{PhysRevLett.83.1359,Smirnov2001CRITICALEF}.

The field $\mathcal{E}$ is only formally defined by \eqref{def:edge} because it cannot be evaluated pointwise, as is customary in (Euclidean) quantum field theory. However,
it is shown in \cite{camia2021scalar} that $\mathcal{E}$ has well-defined $n$-point functions and behaves like a conformal primary field with dimensions $(1/3,1/3)$. The constant $\hat{c}$ and the multiplicative factor $\frac{\hat{c}}{\sqrt\lambda}$ are chosen so that the field is canonically normalized, that is,
\begin{align} \label{eq:canonical_edge}
    \Braket{{\mathcal E}(z_1){\mathcal E}(z_2)}_{\mathbb C} = |z_1-z_2|^{-4/3}.
\end{align}

More generally, the $n$-point functions of the edge operator can be expressed as
\begin{align} \label{eq:n-point-function}
    \langle {\mathcal E}(z_1) \ldots {\mathcal E}(z_n) \rangle_D = \frac{\hat{c}^n}{\lambda^{n/2}} \sum_{(S_1,\ldots,S_r) \in \mathcal{S}} \lambda^r \alpha^{S_1}_D\ldots\alpha^{S_r}_D,
\end{align}
where $\mathcal{S} = \mathcal{S}(z_1,\ldots,z_n)$ denotes the set of all partitions of $\{z_1,\ldots,z_n\}$ such that each element $S_l$ of $(S_1,\ldots,S_r) \in \mathcal{S}$ has cardinality $|S_l| \geq 2$ and, for any subset $S_l=\{z_{j_1},\ldots,z_{j_k}\}$ of $\{z_1,\ldots,z_n\}$, 
\begin{align}
    \alpha^{S_l}_D \equiv \alpha_D^{z_{j_1},\ldots,z_{j_k}} := \lim_{\varepsilon \to 0} \varepsilon^{-2k/3} \mu_D^{\text{loop}}\big(\ell \cap B_{\varepsilon}(z_{j_m}) \neq \emptyset, m=1,\ldots,k \big).
\end{align}

It is shown in \cite{camia2021scalar} that, if $D \subseteq{\mathbb C}$ is either the complex plane $\mathbb C$ or the upper-half plane $\mathbb H$ or any domain conformally equivalent to $\mathbb H$, for any collection of distinct points $z_1,\ldots,z_k \in D$ with $k \geq 2$, the above limit exists and has the following property.
If $D'$ is a domain conformally equivalent to $D$ and $f:D \to D'$ is a conformal map, then
\begin{equation} \label{eq:conf-cov-mu}
    \alpha_{D'}^{f(z_1),\ldots,f(z_k)} = \left( \prod_{j=1}^k |f'(z_k)|^{-2/3} \right) \alpha_D^{z_1,\ldots,z_k}.
\end{equation}
This, combined with \eqref{eq:n-point-function}, implies that
\begin{equation} \label{eq:conf-cov}
    \Braket{\mathcal{E}(f(z_1)) \ldots \mathcal{E}(f(z_n))}_{D'} = \left( \prod_{j=1}^k |f'(z_k)|^{-2/3} \right) \Braket{\mathcal{E}(z_1) \ldots \mathcal{E}(z_n)}_{D}.
\end{equation}

\medskip

\subsection{Summary of the main results and structure of the paper} \label{sec:main-results}

\begin{enumerate}
    \item[Section~\ref{sec:bulk-stress}.] We provide integral expressions of the bulk stress-energy tensor $T$ that can be used to compute certain correlation functions involving $T$. We verify that the Ward identities involving these correlation functions are satisfied, confirming the validity of our expressions for $T$. 
    
    \item[Section~\ref{sec:boundary-edge}.] We study a boundary version of the edge counting operator $\mathcal{E}$ \eqref{def:edge}, namely
    \begin{align}
    \begin{split} \label{def:boundary-edge-normalized}
        {\mathcal B}(\zeta) := \frac{1}{\sqrt\lambda} \lim_{\varepsilon \to 0} \varepsilon^{-2} E_{\varepsilon}(\zeta),
    \end{split}
    \end{align}
    where $\zeta \in \partial D$. We show that, when $D$ is the upper half-plane $\mathbb{H}$ and $x_1,x_2 \in \mathbb{R}$,
    \begin{align}
        \Braket{{\mathcal B}(x_1){\mathcal B}(x_2)}_{\mathbb H} = |x_1-x_2|^{-4}.
    \end{align}
    We identify a new operator, formally
    \begin{align}
    \begin{split}
        {\mathcal T}(\zeta) := \sqrt{\lambda }{\mathcal B}(\zeta) = \lim_{\varepsilon \to 0} \varepsilon^{-2} E_{\varepsilon}(\zeta),
    \end{split}
    \end{align}
    whose behavior is consistent with the role of boundary stress-energy tensor. We derive part of the operator product expansion of $\mathcal{T} \times \mathcal{T}$ \eqref{TT} and check the Ward identity \eqref{eq:Ward}.
    
    \item[Section~\ref{sec:domain-perturbations_bc-ops}.] We link the operator $\mathcal{T}$ to local deformations in the boundary of the domain and to the insertion of a pair of operators that generate a Brownian excursion in the domain (with the insertion points as starting and ending points of the excursion) and appear to behave like \emph{boundary condition changing operators}.
\end{enumerate}

\section{The bulk stress-energy tensor} \label{sec:bulk-stress}

In this section we show how to express the bulk stress-energy tensor $T$ in the BLS in terms of the edge counting operator $\mathcal{E}$ and in terms of the vertex layering operator $\mathcal{O}$.
From these expressions, we compute two examples of correlation functions of primary fields with $T$ and show that they satisfy the conformal Ward identities, confirming the validity of our expressions for $T$.
The strategy of this section is inspired by \cite{Doyon_2006,Gamsa_2006}. In particular, the techniques of \cite{Gamsa_2006}, where the authors derive the central charge of the $O(n \to 0)$ model, have been used in \cite{camia2021scalar} to determine the central charge of the BLS.

\subsection{Identification}

The holomorphic and anti-holomorphic components of the stress-energy tensor can be understood as the generators of conformal transformations.
The holomorphic component of the stress-energy tensor in a two-dimensional CFT can generally be understood as the level-2 descendant of the identity operator
\begin{align} \label{def:L2T}
    (L_{-2} \mathbb{1})(z) \equiv T(z).
\end{align}
The anti-holomorphic component $\bar{T}$ is equivalently given by $\bar{L}_{-2}$. Their conformal dimensions are $(2,0)$ and $(0,2)$ for the holomorphic and anti-holomorphic component, respectively.

\begin{figure}[t]
    \centering
    \begin{subfigure}[b]{0.4\textwidth}
        \includegraphics[width=\textwidth]{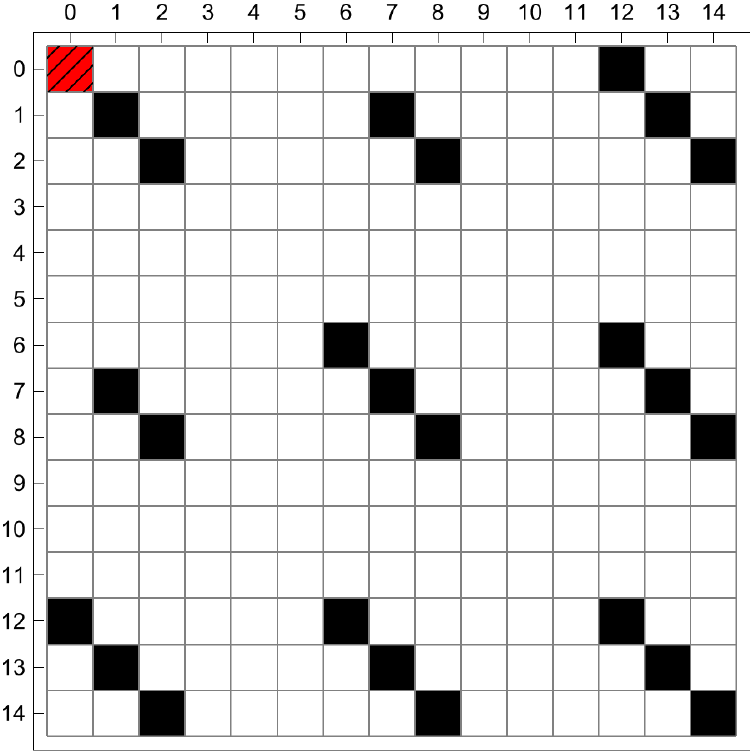}
        \caption{Non-zero $C_{\mathcal{E}\mathcal{E}}^{(p,p')}$}
        \label{VVEE_blocks}
    \end{subfigure} ~
    \begin{subfigure}[b]{0.4\textwidth}
        \includegraphics[width=\textwidth]{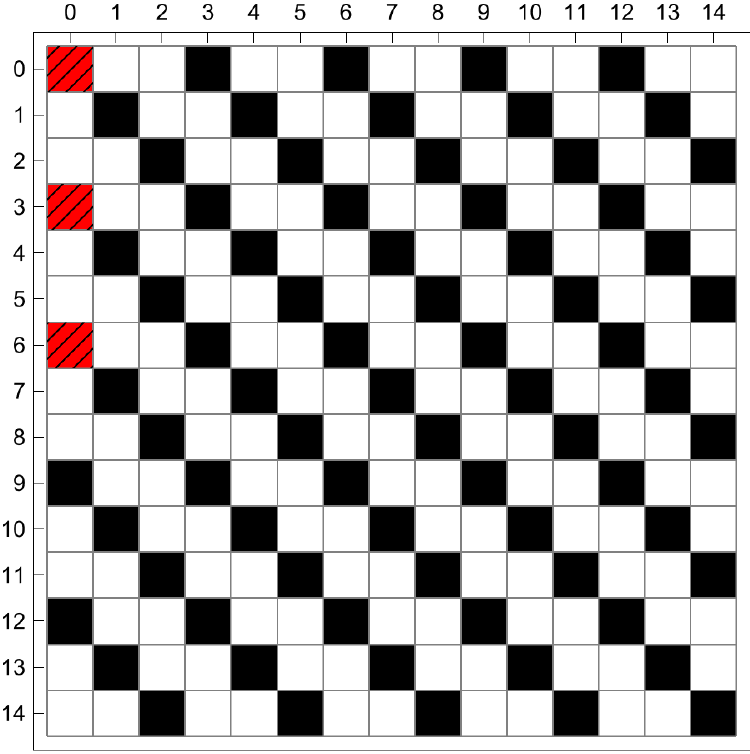}
        \caption{Non-zero $C_{\mathcal{O}_\beta \mathcal{O}_{-\beta}}^{(p,p')}$}
        \label{VVVV_blocks}
    \end{subfigure}
    \caption{The non-zero three-point function coefficients are shown. Rows and columns label $(p,p')$ for primary operators with dimensions $(p/3, p'/3)$.  Left: between two edge-counting operators. Right: between two vertex operators.
    The operators giving contributions to $T$, corresponding to $\mathbb{1}$ (left) and $\mathbb{1}, J, W$ (right) are marked in red (hatched).}
    \label{checkerboard}
\end{figure}

\begin{figure}[t]
    \centering
    \includegraphics[width=.75\textwidth]{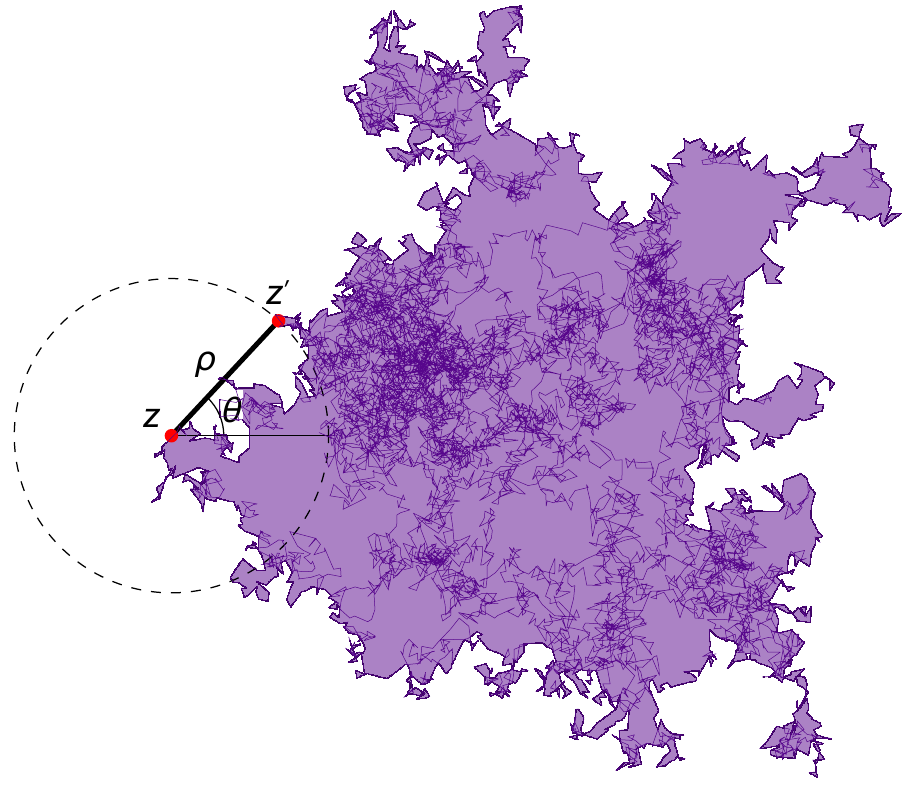}
    \caption{Every loop whose outer boundary comes close to the point $z$ and intersects the dashed circle somewhere contributes to $T(z)$ in \eqref{def:T}.
    The red points represent the insertion points of the edge-counting operators $\mathcal{E}$ at $z$ and $z' = z + \rho e^{i \theta}$.}
    \label{fig:slit}
\end{figure}

Our strategy is to identify the stress-energy tensor starting from the operator product expansion (OPE) of two primary operators $\mathcal{A}_1$ and $\mathcal{A}_2$ (see Section 6.6.3 of \cite{DiFrancesco:639405})
\begin{align}
\begin{split} \label{OPE}
    \mathcal{A}_1(z+\epsilon) \times \mathcal{A}_2(z) = \sum_\mathcal{P} \sum_{\{k,\bar{k}\}} & C^\mathcal{P}_{12} \beta^{\mathcal{P}\{k\}}_{12} \bar{\beta}^{\mathcal{P}\{\bar{k}\}}_{12} \epsilon^{\Delta_\mathcal{P} - \Delta_1 - \Delta_2 + K} \bar{\epsilon}^{\bar{\Delta}_\mathcal{P} - \bar{\Delta}_1 - \bar{\Delta}_2 + \bar{K}} \\
    & \cdot L_{-k_1}\ldots L_{-k_N} \bar{L}_{-\bar{k}_1}\ldots \bar{L}_{-\bar{k}_{\bar{N}}} \mathcal{P}(z),
\end{split}
\end{align}
where the outer sum runs over all primary operators $\mathcal{P}$, and the inner sum is over all collections of indices (the descendant levels)
$k_i, \bar{k}_i$ with 
$K=\sum_{k_i \in \{k\}} k_i$ and $\bar{K}=\sum_{\bar{k}_i \in \{\bar{k}\}} \bar{k}_i$.
$\beta_{1 2}^{\mathcal{P}\{k\}}, \bar{\beta}_{1 2}^{\mathcal{P}\{\bar{k}\}}$ are numerical coefficients fully determined by the Virasoro algebra (they depend on the central charge and the conformal dimensions of the operators involved),
and $C_{1 2}^\mathcal{P}$ are three-point function coefficients of the theory.
If we choose $\mathcal{A}_1$ and $\mathcal{A}_2$ so that \eqref{OPE} contains the identity operator then the stress tensor will appear as a descendent. In particular, the identity is contained in the OPE
\begin{align}
\begin{split} \label{OPE-EE}
    \mathcal{E}(z+\epsilon) \times \mathcal{E}(z) & = |\epsilon|^{-4/3} \left( \mathbb{1} +
    \epsilon^2 \frac{2\Delta_{\mathcal{E}}}{c}T(z) + \ldots
    \right) \\
    & = |\epsilon|^{-4/3} \left( \mathbb{1} +
    \epsilon^2 \frac{1/3}{\lambda}T(z) + \ldots
    \right),
\end{split}
\end{align}
where $\Delta_{\mathcal{E}}=1/3$ is the conformal dimension of $\mathcal{E}$ and $c=2\lambda$ is the central charge, and where higher descendants, as well as other conformal blocks and the anti-holomorphic components, have been omitted. The term containing $T(z)$ appears at order $O(\epsilon^{-2/3 + 2} \bar{\epsilon}^{-2/3})$ in \eqref{OPE-EE}, and no other term appears at the same order. (The anti-holomorphic component $\bar{T}$ would appear at order $O(\epsilon^{-2/3} \bar{\epsilon}^{-2/3+2})$.) This can be deduced from the analysis carried out in \cite{Camia_2020}, as shown in Figure \ref{VVEE_blocks}.
Thus one can identify
\begin{align} \label{def:T}
    T(z) = \frac{\lambda}{1/3} \lim_{\rho \to 0} \rho^{-2/3}
    \frac{1}{2\pi} \int_0^{2\pi} d \theta ~ e^{-2 i \theta}
    \mathcal{E}(z+\rho e^{i \theta}) \mathcal{E}(z),
\end{align}
since the integral in \eqref{def:T} singles out the term of the expansion in the right hand side of \eqref{OPE-EE} that corresponds to the stress-energy tensor. 
As shown in Figure \ref{fig:slit}, the loops that contribute to the integral in \eqref{def:T} are those that come close to $z$ and that intersect the circle of radius $\rho$ centered at $z$. This provides a geometric interpretation for the stress-energy tensor.
$\bar{T}$ is obtained by replacing $e^{-2i \theta}$ with $e^{2i \theta}$.

As a consequence of conformal invariance, correlation functions involving the stress-energy tensor in the bulk obey the conformal Ward identities
\begin{align} \label{eq:wardbulk}
    \Braket{ T(z) \prod_{j=1}^N \mathcal{A}_j(z_j) }_\mathbb{C} = \sum_{j=1}^N \left(
    \frac{\Delta_j}{(z-z_j)^2} + \frac{1}{z-z_j}\partial_{z_j}
    \right) \Braket{\prod_{j=1}^N \mathcal{A}_j(z_j)}_\mathbb{C},
\end{align}
where $\mathcal{A}_j$ denotes a generic primary operator with conformal dimension $\Delta_j$. In the next section, we test \eqref{def:T} using \eqref{eq:wardbulk}.

\subsection{Correlation functions}

We now use the full-plane mixed four-point function $\Braket{\mathcal{O}_{\beta}(z_1) \mathcal{O}_{-\beta}(z_2) {\mathcal E}(z_3) {\mathcal E}(z_4)}_\mathbb{C}$ to compute $\braket{T(z)\mathcal{O}_\beta(z_1)\mathcal{O}_{-\beta}(z_2)}_\mathbb{C}$ and $\braket{T(z)\mathcal{E}(z_1)\mathcal{E}(z_2)}_\mathbb{C}$, which can then be compared to the Ward identities \eqref{eq:wardbulk}. (See Section \ref{sec:definitions} for the definition of $\mathcal{O}_\beta(z)$.) The four-point function was computed in \cite{camia2021scalar} and can be written as
\begin{align}
\begin{split} \label{eq:OOEE}
    & \Braket{\mathcal{O}_{\beta}(z_1) \mathcal{O}_{-\beta}(z_2) {\mathcal E}(z_3) {\mathcal E}(z_4)}_\mathbb{C} \\
    & = \Braket{\mathcal{O}_{\beta}(z_1) \mathcal{O}_{-\beta}(z_2)}_\mathbb{C} \Big[ \hat{\alpha}_\mathbb{C}^{z_3,z_4} - (1-\cos\beta) \hat{\alpha}^{z_3,z_4}_{z_1|z_2;\mathbb{C}} + \lambda (1-\cos\beta)^2 \hat{\alpha}^{z_3}_{z_1|z_2;\mathbb{C}} \hat{\alpha}^{z_4}_{z_1|z_2;\mathbb{C}} \Big],
\end{split}
\end{align}
where, letting $z_{jk}=z_j-z_k$, 
\begin{align}
\begin{split} \label{eq:alphas}
        \Braket{\mathcal{O}_{\beta}(z_1) \mathcal{O}_{-\beta}(z_2)}_\mathbb{C} &= |z_{12}|^{-4\Delta(\beta)} \\
        \Braket{\mathcal{E}(z_3) \mathcal{E}(z_4)}_\mathbb{C} = \hat\alpha^{z_3,z_4}_\mathbb{C} &= |z_{34}|^{-4/3} \\
        \hat{\alpha}^{z_3}_{z_1|z_2;{\mathbb C}} & = \frac{ 2^{7/6} \pi}{3^{1/4} \sqrt{5} \, \Gamma(1/6)\Gamma(4/3)} \left| \frac{z_{12}}{z_{13}z_{23}} \right|^{2/3} \\
        \hat\alpha^{z_3,z_4}_{z_1|z_2;\mathbb{C}} & = \frac{|z_{34}|^{-4/3}-Z_{\text{twist}}}{2},
\end{split}
\end{align}
and
\begin{align}
\begin{split} \label{ztwist}
    & Z_{\text{twist}} = Z_{\text{twist}}(z_1,z_2;z_3,z_4) \\
    &= \left| \frac{z_{13} z_{24} }{ z_{34}^2 z_{23} z_{14} } \right|^{2/3} \left[
    \left| {}_2 F_1\left( -\frac{2}{3}, \frac{1}{3}; \frac{2}{3}, x \right)  \right|^2 - \frac{4\Gamma\left( \frac{2}{3} \right)^6}{\Gamma\left( \frac{4}{3} \right)^2 \Gamma\left( \frac{1}{3} \right)^4} |x|^{2/3} \left| {}_2 F_1\left( -\frac{1}{3}, \frac{2}{3}; \frac{4}{3}, x \right) \right|^2
    \right]
\end{split}
\end{align}
corresponds to Eq.\ (52) of \cite{Simmons_2009} with $x=\frac{z_{12}z_{34}}{z_{13}z_{24}}$.

We can then apply \eqref{def:T} to \eqref{eq:OOEE} using \eqref{eq:alphas} by expanding $\hat{\alpha}^{z+\rho e^{i \theta}}_{z_1|z_2;\mathbb{C}}$ in powers of $\rho$. Observe that the terms independent of $\theta$ (e.g., $\hat\alpha_{\mathbb{C}}^{z,z+\rho e^{i \theta}}=\rho^{-3/4}$) do not contribute because of the integral $\int_0^{2\pi} d \theta ~ e^{-2i \theta}$, while the terms containing powers of $\rho$ greater than $2/3$ give zero in the limit $\rho \to 0$.
This gives
\begin{align}
\begin{split} \label{eq:TOO-Ztwist}
    & \Braket{T(z) \mathcal{O}_\beta(z_1) \mathcal{O}_{-\beta}(z_2)}_{\mathbb{C}} \\
    &= \frac{\lambda}{1/3}
    \lim_{\rho \to 0} \rho^{-2/3}
    \frac{1}{2\pi} \int_{0}^{2\pi} d \theta ~ e^{-2i \theta}
    \Braket{\mathcal{E}(z+\rho e^{i \theta}) \mathcal{E}(z) \mathcal{O}_\beta(z_1) \mathcal{O}_{-\beta}(z_2)}_{\mathbb{C}} \\
    &= \frac{3\lambda(1-\cos\beta)}{2} |z_{12}|^{-4\Delta(\beta)}
    \lim_{\rho \to 0} \rho^{-2/3}
    \frac{1}{2\pi} \int_{0}^{2\pi} d \theta ~ e^{-2i \theta}
    Z_{\text{twist}}(z_1,z_2;z,z+\rho e^{i \theta}) \, .
\end{split}
\end{align}
Using the fact that $\rho^{-2/3} Z_{\text{twist}}(z_1,z_2;z,z+\rho e^{i \theta})$ is analytic and expanding it in powers of $\rho$, performing the integral, and sending $\rho \to 0$, we obtain
\begin{align}
\begin{split} \label{eq:TOO}
    \Braket{T(z) \mathcal{O}_\beta(z_1) \mathcal{O}_{-\beta}(z_2)}_{\mathbb{C}} = \Delta(\beta) \, \frac{ (z_1-z_2)^{2-2\Delta(\beta)} (\bar{z}_1-\bar{z}_2)^{-2\Delta(\beta)} }
    { (z-z_1)^2 (z-z_2)^2 } \, ,
\end{split}
\end{align}
which corresponds to the Ward identity \eqref{eq:wardbulk} for $N=2, \mathcal{A}_{1,2} = \mathcal{O}_{\pm \beta}$, and $\Delta_j = \Delta(\beta)$.

In a very similar manner, one can obtain $\Braket{T(z) \mathcal{E}(z_3) \mathcal{E}(z_4)}_\mathbb{C}$ using the fact that the identity block is also contained in the OPE of $\mathcal{O}_{\beta}(z+\epsilon) \times \mathcal{O}_{-\beta}(z)$. The term containing $T(z)$ appears at order $O(\epsilon^{-2 \Delta(b) + 2} \bar{\epsilon}^{-2 \Delta(b)})$ (the anti-holomorphic component $\bar{T}$ appears at order $O(\epsilon^{-2 \Delta(\beta)} \bar{\epsilon}^{-2 \Delta(\beta)+2})$), but in this case there are other contributions to the OPE at the same order. The relevant operators are conformal primaries with dimensions $(1,0)$ and $(2,0)$, and we denote them $J$ and $W$, respectively. They are shown in Figure \ref{VVVV_blocks}, which is obtained using results from \cite{Camia_2020}. More precisely, in the case of two layering vertex operators, \eqref{OPE} gives
\begin{align}
\begin{split} \label{OPEapplied}
    &\mathcal{O}_\beta(z+\epsilon) \times \mathcal{O}_{-\beta}(z) \\
    &= |\epsilon|^{-4\Delta(\beta)} \epsilon^2 \left( \beta_{\mathcal{O}_\beta \mathcal{O}_{-\beta}}^{\mathbb{1} \{2\}} (L_{-2} \mathbb{1})(z) + \beta_{\mathcal{O}_\beta \mathcal{O}_{-\beta}}^{J\{1\}} C_{\mathcal{O}_\beta \mathcal{O}_{-\beta}}^J (L_{-1} J)(z) + + C_{\mathcal{O}_\beta \mathcal{O}_{-\beta}}^W W(z) \right) + \ldots \\
    &= |\epsilon|^{-4\Delta(\beta)} \epsilon^2 \left(
    \frac{2 \Delta(\beta)}{c}T(z) + \frac{1}{2} C_{\mathcal{O}_\beta \mathcal{O}_{-\beta}}^J \partial_z J(z) + C_{\mathcal{O}_\beta \mathcal{O}_{-\beta}}^W W(z)
    \right) + \ldots,
\end{split}
\end{align}
where higher descendants, other conformal blocks, as well as the anti-holomorphic components, have been omitted. The coefficients $\beta_{\mathcal{O}_\beta \mathcal{O}_{-\beta}}^{\mathbb{1} \{2\}}$ and $\beta_{\mathcal{O}_\beta \mathcal{O}_{-\beta}}^{J\{1\}}$ are determined by the Virasoro algebra and are given by $2 \Delta(\beta) / c$ and $1/2$, respectively \cite{DiFrancesco:639405}; the central charge of the BLS is $c=2\lambda$, $(L_{-1} J)(z) = \partial_z J(z)$, and $C_{\mathcal{O}_\beta \mathcal{O}_{-\beta}}^{\mathbb{1}} = 1$ because of the (canonical) normalization of $\mathcal{O}_{\beta}$. Finally, the three-point function coefficients $C_{\mathcal{O}_\beta \mathcal{O}_{-\beta}}^{J}$ and $C_{\mathcal{O}_\beta \mathcal{O}_{-\beta}}^{W}$ can be computed\footnote{In \cite{Camia_2020} we found that $C_{\mathcal{O}_{\beta_1} \mathcal{O}_{\beta_2} }^J = 0$ for generic $\beta_1, \beta_2$. In the special case $\beta_1 + \beta_2 = 0 \mod 2 \pi$ the coefficient is given in \eqref{eq:threepcoeffs}.} as explained in \cite{Camia_2020}, which gives
\begin{align}
\begin{split} \label{eq:threepcoeffs}
    \left( C_{\mathcal{O}_\beta \mathcal{O}_{-\beta }}^J \right)^2 &= \frac{\lambda}{10} \sin^2\beta \\
    \left( C_{\mathcal{O}_\beta \mathcal{O}_{-\beta }}^W \right)^2 &= \frac{1}{2} \left( C_{\mathcal{O}_\beta \mathcal{O}_{-\beta }}^J \right)^4.
\end{split}
\end{align}
Based on these considerations, one can write
\begin{align}
\begin{split} \label{def:T-alternative}
    T(z) & = \frac{\lambda}{\Delta(\beta)} \bigg[ \lim_{\rho \to 0} \rho^{4\Delta(\beta)-2}
    \frac{1}{2\pi} \int_0^{2\pi} d \theta ~ e^{-2i \theta}
    \mathcal{O}_\beta(z+\rho e^{i \theta}) \mathcal{O}_{-\beta}(z) \\
    & \qquad - \frac{1}{2} C_{\mathcal{O}_\beta \mathcal{O}_{-\beta}}^J \partial_z J(z) - C_{\mathcal{O}_\beta \mathcal{O}_{-\beta}}^W W(z)
    \bigg].
\end{split}
\end{align}

Applying \eqref{def:T-alternative}, \eqref{eq:OOEE} and \eqref{eq:alphas}, and using the fact that $\braket{J \mathcal{E} \mathcal{E}}_{\mathbb{C}} = \braket{ W \mathcal{E} \mathcal{E}}_{\mathbb{C}} = 0$, as can be seen from Figure \ref{VVEE_blocks}, we obtain
\begin{align}
\begin{split} \label{eq:calculation}
    &\Braket{T(z) {\mathcal E}(z_3) {\mathcal E}(z_4)}_{\mathbb{C}} \\
    & \quad = \frac{\lambda}{\Delta(\beta)}
    \lim_{\rho \to 0} \rho^{4\Delta(\beta)-2}
    \frac{1}{2\pi} \int_{0}^{2\pi} d \theta ~ e^{-2i \theta}
    \Braket{\mathcal{O}_{\beta}(z+\rho e^{i \theta}) \mathcal{O}_{-\beta}(z) {\mathcal E}(z_3) {\mathcal E}(z_4)}_{\mathbb{C}} \\
    & \quad = - \frac{\lambda(1-\cos\beta)}{\Delta(\beta)}
    \lim_{\rho \to 0} \rho^{-2}
    \frac{1}{2\pi} \int_{0}^{2\pi} d \theta ~ e^{-2i \theta} \,  \hat\alpha^{z_3,z_4}_{z|z+\rho e^{i\theta};\mathbb{C}} \\
    & \quad = \frac{\lambda(1-\cos\beta)}{2\Delta(\beta)}
    \lim_{\rho \to 0} \rho^{-2}
    \frac{1}{2\pi} \int_{0}^{2\pi} d \theta ~ e^{-2i \theta} \, Z_{\text{twist}}(z,z+\rho e^{i\theta};z_3,z_4) \\
    & \quad =  \frac{1}{3} \frac{(z_3-z_4)^{4/3}}{(z-z_3)^2 (z-z_4)^2 (\bar{z}_3-\bar{z}_4)^{2/3}} \, .
\end{split}
\end{align}
This is the correct form of the conformal Ward identity in the bulk, i.e.\ \eqref{eq:wardbulk}, with $N=2, \mathcal{A}_j = \mathcal{E}$ and $\Delta_j = 1/3$.

We point out that using \eqref{def:T-alternative} in
\begin{align}
    \Braket{\mathcal{O}_{\tilde\beta}(z+\rho e^{i\theta}) \mathcal{O}_{-\tilde\beta}(z) \mathcal{O}_{\beta}(z_1) \mathcal{O}_{-\beta}(z_2)}_{\mathbb{C}},
\end{align}
which has been derived in \cite{Camia_2020}, and performing a similar analysis that takes into account the contributions from $J$ and $W$ from \eqref{OPEapplied}, leads again to \eqref{eq:TOO}, as expected. 

To conclude this section, we compute the bulk stress-energy tensor two-point function, which is fixed by conformal invariance to be $\Braket{T(z)T(z')}_\mathbb{C} = c/2/(z-z')^4$. We apply \eqref{OPE-EE} again to \eqref{eq:calculation} by setting $z_3 = z' + \rho e^{i \theta}, \bar{z}_3 = \bar{z}' + \rho e^{-i \theta}, z_4 = z', \bar{z}_4 = \bar{z}'$ in \eqref{eq:calculation} and obtain
\begin{align}
\begin{split}
    &\Braket{T(z)T(z')}_\mathbb{C} \\
    &= \frac{\lambda}{1/3} \lim_{\rho \to 0} \rho^{-2/3} \frac{1}{2 \pi} \int_0^{2\pi} d \theta ~ e^{-2 i \theta} \frac{1}{3} \frac{(\rho e^{i \theta})^{4/3}}{(z-z'-\rho e^{i\theta})^2 (z-z')^2 (\rho e^{-i\theta})^{2/3}} \\
    &= \frac{\lambda}{(z-z')^4}.
\end{split}
\end{align}
This is indeed the correct form of the two-point function of stress-energy tensors in a theory with central charge $c = 2 \lambda$.

\section{The boundary stress-energy tensor} \label{sec:boundary-edge}

In this section we introduce a boundary version of the edge operator discussed in Section \ref{sec:edge} and study some of its properties. In particular, we show that it is essentially the boundary stress-energy tensor (up to a factor of $\sqrt{\lambda}$).

\begin{figure}[t]
    \centering
    \includegraphics[width=.8\textwidth]{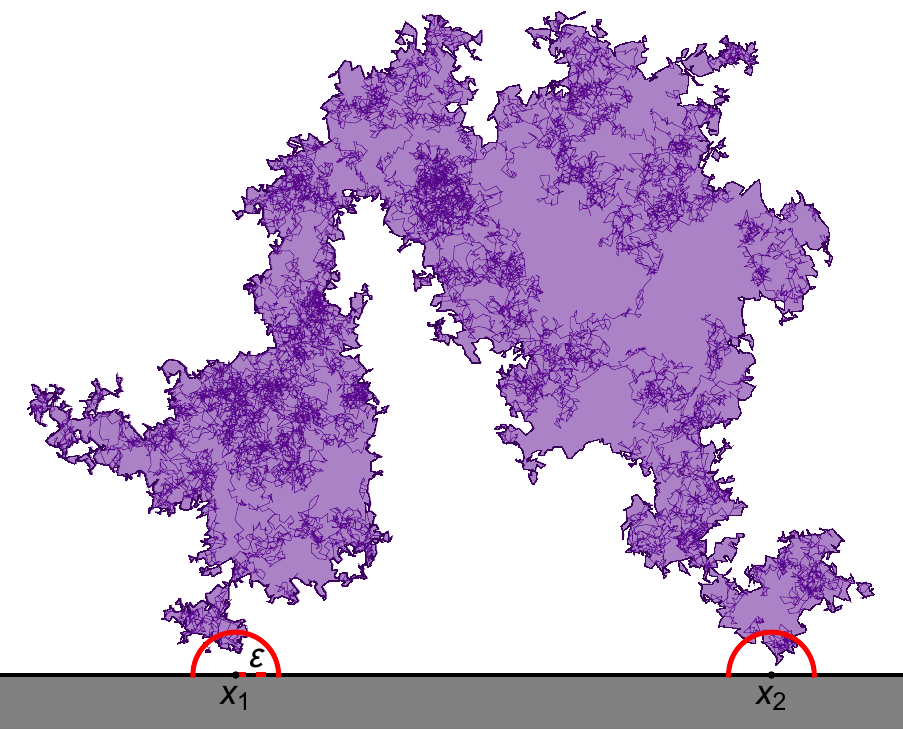}
    \caption{A Brownian loop and its outer boundary in the upper-half plane. Such a loop would contribute to the two-point function of boundary edge counting operators inserted at $x_{1}$ and $x_2$ because the loop comes within distance $\varepsilon$ of both points.}
    \label{edge_edge}
\end{figure}

\subsection{The boundary edge counting operator and its correlation functions}

The same analysis briefly discussed in the Section \ref{sec:edge} can be carried out for a point $\zeta$ on the boundary $\partial D$ of a domain $D$, where $D$ is either a finite domain with a sufficiently smooth boundary or the upper half plane $\mathbb{H}$.
In this case, using the connection with critical percolation and CLE$_6$, the relevant probability is that of a three-arm event at the boundary, which gives an exponent $2$ instead of $2/3$. This is consistent with the well-known fact that in a renormalizable quantum field theory operators at the boundary require a different renormalization from those in the bulk \cite{CARDY2006333} (producing, in general, a different set of conformal dimensions) and leads to the following (formal) definition of \emph{boundary edge counting operator}
\begin{align}
\begin{split} \label{def:boundary-edge}
    {\mathcal B}(\zeta) := \frac{\hat{c}_b}{\sqrt\lambda} \lim_{\varepsilon \to 0} \varepsilon^{-2} E_{\varepsilon}(\zeta),
\end{split}
\end{align}
where $\hat{c}_b$ is chosen so that ${\mathcal B}$ is canonically normalized when $D=\mathbb{H}$, that is, for points $x_1,x_2$ on the real line,
\begin{equation} \label{eq:canonical_boundary}
\Braket{{\mathcal B}(x_1){\mathcal B}(x_2)}_{\mathbb H} = |x_1-x_2|^{-4}.
\end{equation}
At the end of Section \ref{sec:boundary-stress-energy-tensor}, we will see that $\hat{c}_b=1$.

Definition \eqref{def:boundary-edge} is convenient because it uses the edge counting operator analyzed in \cite{camia2021scalar}, but we could have equivalently defined $\mathcal{B}$ by counting the loops crossing an infinitesimal slit attached to the boundary of the domain. This alternative definition, reminiscent of the definitions of the operators discussed in \cite{FRIEDRICH2002947,Doyon_2006}, would not change our conclusions.

The existence of the $n$-point functions of the field ${\mathcal B}$ follows from the same arguments used in \cite{camia2021scalar}. First of all, we note that the proof that the limit in \eqref{eq:E-correlations} exists  extends trivially to the case in which (some of) the points $z_j$ are moved to the boundary of the domain (see Lemma A.1 of \cite{camia2021scalar}). This means that
\begin{align}
    \Braket{ E_{\varepsilon}(x_1) \ldots E_{\varepsilon}(x_n) }_{\mathbb{H}} = \lim_{\delta \to 0}\Braket{ E^{\delta}_{\varepsilon}(x_1) \ldots E^{\delta}_{\varepsilon}(x_n) }_{\mathbb{H}}
\end{align}
exists for all $x_1,\ldots,x_n \in \mathbb{R}$ with $n \geq 2$, which allows us to state the following result.
\begin{proposition} \label{prop:n-point-functions}
For any collection of distinct points $x_1, \ldots, x_n$ on the real line, with $n \geq 2$, the following limit exists:
\begin{align} \label{eq:lim-existence}
\begin{split}
\Braket{\mathcal{B}(x_1) \ldots \mathcal{B}(x_n)}_{\mathbb{H}}
\quad := \frac{\hat{c}_b^n}{\lambda^{n/2}} \lim_{\varepsilon \to 0} \varepsilon^{-2n} \Braket{ E_{\varepsilon}(x_1) \ldots E_{\varepsilon}(x_n) }_{\mathbb{H}}.
\end{split}
\end{align}
Moreover
\begin{equation}
    \Braket{\mathcal{B}(x_1) \ldots \mathcal{B}(x_n)}_{\mathbb{H}} = \frac{\hat{c}_b^n}{\lambda^{n/2}} \sum_{(S_1,\ldots,S_r) \in \mathcal{S}} \lambda^r \beta^{S_1}\ldots\beta^{S_r},
\end{equation}
where $\mathcal{S} = \mathcal{S}(x_1,\ldots,x_n)$ denotes the set of all partitions of $\{x_1,\ldots,x_n\}$ such that each element $S_l$ of $(S_1,\ldots,S_r) \in \mathcal{S}$ has cardinality $|S_l| \geq 2$ and, for any subset $S_l=\{x_{j_1},\ldots,x_{j_k}\}$ of $\{x_1,\ldots,x_n\}$, 
\begin{align}
    \beta^{S_l} \equiv \beta^{x_{j_1},\ldots,x_{j_k}} := \lim_{\varepsilon \to 0} \varepsilon^{-2k} \mu^{\text{loop}}_{\mathbb{H}}\left(\ell \cap B_{\varepsilon}(x_{j_m}), \, m=1,\ldots,k \right).
\end{align}
\end{proposition}

\noindent{\bf Proof.} The proposition is analogous to Theorem 2.3 of \cite{camia2021scalar} and its proof follows directly from the arguments used in the proofs of Lemmas 2.1 and 2.2 of \cite{camia2021scalar}. \qed

\medskip

The two-point function $\Braket{{\mathcal B}(x_1){\mathcal B}(x_2)}_{\mathbb H}$ can be calculated exactly, as follows. For any $s>0$, the scale invariance of the BLS implies that
\begin{align}
\begin{split} \label{eq:scaling-two-point-func}
    \Braket{{\mathcal B}(sx_1){\mathcal B}(sx_2)}_{\mathbb H} &= \frac{\hat{c}_b^2}{\lambda} \lim_{\varepsilon \to 0} \varepsilon^{-4} \Braket{ E_{\varepsilon}(sx_1) E_{\varepsilon}(sx_2) }_{\mathbb{H}} \\
    &= s^{-4} \, \frac{\hat{c}_b^2}{\lambda} \lim_{\varepsilon \to 0} \left(\frac{\varepsilon}{s}\right)^{-4} \Braket{ E_{\varepsilon/s}(x_1) E_{\varepsilon/s}(x_2) }_{\mathbb{H}}\\
    &= s^{-4} \Braket{{\mathcal B}(x_1){\mathcal B}(x_2)}_{\mathbb H}.
\end{split}
\end{align}
Since $\Braket{{\mathcal B}(x_1){\mathcal B}(x_2)}_{\mathbb H}$ can only depend on $|x_1-x_2|$, this means that
\begin{align}
    \Braket{{\mathcal B}(x_1){\mathcal B}(x_2)}_{\mathbb H} \sim |x_1-x_2|^{-4},
\end{align}
and we can choose $\hat{c}_b$ so that the proportionality constant is $1$. (As mentioned above, it will turn out that $\hat{c}_b=1$.)

\subsection{Identification of the boundary stress-energy tensor} \label{sec:boundary-stress-energy-tensor}

We now present the main results of this section, which allow us to identify $\sqrt{\lambda}\mathcal{B}$ with the boundary stress-energy tensor of the theory. To this end, we recall the formal definition of \emph{layering vertex operator} in the upper-half plane \cite{Camia_2020}
\begin{align} \label{def:tildeO}
    \tilde{\mathcal{O}}_{\beta}(z) := \lim_{\delta \to 0} (\hat{c}_{\mathbb{H}}\delta)^{-2\Delta(\beta)} \exp{\Big(i \beta \sum_{\stackrel{\ell \in \mathcal{L}^{\delta} }{z \in \bar\ell}} \sigma_{\ell} \Big)},
\end{align}
where $\Delta(\beta) = \bar{\Delta}(\beta) = \frac{\lambda}{10}(1-\cos\beta)$, $\sigma_{\ell}$ is a symmetric $(\pm 1)$-valued random variable associated to loop $\ell$, and $\mathcal{L}^{\delta}$ is a BLS from which all loops of diameter smaller than $\delta$ have been removed, $z \in \bar\ell$ means that $z$ is in the interior of the domain bounded by $\ell$ (it is disconnected from infinity by $\ell$); $\hat{c}_{\mathbb{H}}$ is chosen so that
\begin{equation} \label{eq:layering-canonical-two-point-func}
\Braket{\tilde{\mathcal{O}}_{\beta}(z)}_{\mathbb H} = |z-\bar{z}|^{-2\Delta(\beta)}.
\end{equation}
Equation \eqref{eq:layering-canonical-two-point-func} is obtained in \cite{Camia_2020}, while the existence and conformal covariance properties of more general $n$-point functions are proved in \cite{Camia_2016}.
Essentially, when a layering operator is inserted at $z$, each loop $\ell$ such that $z \in \bar\ell$ picks up a phase $e^{i \beta\sigma_{\ell}}$. We note that the layering vertex operators are scalar operators.

\begin{proposition} \label{prop:OOB}
For any choice of points $z_1,z_2 \in \mathbb{H}$, we have that
\begin{align}
\begin{split} \label{eq:OOB}
    & \Braket{\mathcal{B}(0) \tilde{\mathcal{O}}_{\beta}(z_1) \tilde{\mathcal{O}}_{-\beta}(z_2) }_{\mathbb{H}} = -\hat{c}_b \sqrt{\lambda} (1-\cos\beta) \Braket{\tilde{\mathcal{O}}_{\beta}(z_1) \tilde{\mathcal{O}}_{-\beta}(z_2)}_{\mathbb{H}} \\
    & \quad \cdot \left\{ \frac{2}{5} \left[ \left(\Im\left(\frac{1}{z_1}\right)\right)^2 + \left(\Im\left(\frac{1}{z_2}\right)\right)^2 \right] - \frac{4}{5} \Im\left(\frac{1}{z_1}\right) \Im\left(\frac{1}{z_2}\right) G(\sigma(z_1,z_2)) \right\}, 
\end{split}
\end{align}
where $\sigma(z_1,z_2) = \frac{|z_1-z_2|^2}{|z_1-\bar{z}_2|^2}$,
\begin{align}
    G(\sigma) = 1 - \sigma \, {}_2 F_1\left(1,\frac{4}{3};\frac{5}{3};1-\sigma \right),
\end{align}
and ${}_2 F_1$ is the hypergeometric function.
\end{proposition}
\noindent{\bf Proof.}
Using (3.6) and (3.7) from \cite{camia2021scalar} applied to the upper half-plane and with $z_3=0$ and the exponent $2/3$ replaced by $2$, we obtain an analog of (3.8) therein, namely
\begin{align}
\begin{split} \label{eq:OOB-proof}
    & \Braket{\tilde{ \mathcal{B}(0) \mathcal{O}}_{\beta}(z_1) \tilde{\mathcal{O}}_{-\beta}(z_2) }_{\mathbb{H}} = - \lambda (1-\cos\beta) \Braket{\tilde{\mathcal{O}}_{\beta}(z_1) \tilde{\mathcal{O}}_{-\beta}(z_2)}_{\mathbb{H}} \\
    &\qquad \cdot \frac{\hat{c}_b}{\sqrt{\lambda}} \lim_{\varepsilon \to 0} \varepsilon^{-2} \mu^{\text{loop}}_{\mathbb{H}}\big( \ell \cap B_{\varepsilon}(0) \neq \emptyset, \ell \text{ separates } z_1,z_2 \big).
\end{split}
\end{align}
In order to evaluate the limit in \eqref{eq:OOB-proof}, we use the fact that
\begin{align}
\begin{split} \label{eq:measures}
    \lim_{\varepsilon \to 0} \varepsilon^{-2} \mu^{\text{loop}}_{\mathbb{H}}\big( \ell \cap B_{\varepsilon}(0) \neq \emptyset, \ell \text{ separates } z_1,z_2 \big) = \mu^{\text{bub}}_{\mathbb{H}}(\ell \text{ separates } z_1,z_2),
\end{split}
\end{align}
where $\mu^{\text{bub}}_{\mathbb{H}}$ is the measure on Brownian bubbles in $\mathbb{H}$, pinned at the origin, defined in \cite{lawler2004brownian}, and the equality follows from Theorem 1 of \cite{lawler2004brownian}.

As explained in Remark 7.1 of \cite{Lawler_2003} and in \cite{beliaev2013some}, $\mu^{\text{bub}}_{\mathbb{H}}$ is closely related to the SLE$_{8/3}$ excursion measure $\mu^{\text{SLE}_{8/3}}_{\mathbb{H}}$; more precisely $\mu^{\text{bub}}_{\mathbb{H}} = \frac{8}{5} \mu^{\text{SLE}_{8/3}}_{\mathbb{H}}$. Therefore we have
\begin{align}
\begin{split} \label{eq:OOB-SLE}
& \Braket{ \mathcal{B}(0) \tilde{\mathcal{O}}_{\beta}(z_1) \tilde{\mathcal{O}}_{-\beta}(z_2) }_{\mathbb{H}} \\
&= - \sqrt{\lambda} \hat{c}_b (1-\cos\beta) \Braket{\tilde{\mathcal{O}}_{\beta}(z_1) \tilde{\mathcal{O}}_{-\beta}(z_2)}_{\mathbb{H}} \frac{8}{5} \mu^{\text{SLE}_{8/3}}_{\mathbb{H}}(\ell \text{ separates } z_1,z_2) \\
&= - \sqrt{\lambda} \hat{c}_b (1-\cos\beta) \Braket{\tilde{\mathcal{O}}_{\beta}(z_1) \tilde{\mathcal{O}}_{-\beta}(z_2)}_{\mathbb{H}} \\
& \qquad \cdot \frac{8}{5} \left( \mu^{\text{SLE}_{8/3}}_{\mathbb{H}}(z_1 \in \bar\ell) + \mu^{\text{SLE}_{8/3}}_{\mathbb{H}}(z_2 \in \bar\ell) -2 \mu^{\text{SLE}_{8/3}}_{\mathbb{H}}(z_1,z_2 \in \bar\ell) \right).
\end{split}
\end{align}
Since $\mu^{\text{SLE}_{8/3}}$ is obtained by multiplying the law of an SLE$_{8/3}$ 
$\varepsilon$-bubble (i.e., an SLE$_{8/3}$ excursion in $\mathbb{H}$ starting at the origin and ending at $\varepsilon$) by $\varepsilon^{-2}$ and passing to the limit, the proof is concluded by using (3.2) and (3.3) from Proposition 1 of \cite{beliaev2013some} to evaluate the three terms in the last line of the last equation. \qed

\medskip

The importance of Proposition \ref{prop:OOB} lies in the fact that it allows us to identify the boundary stress-energy tensor of our CFT. The ingredients for this identification are
\begin{itemize}
    \item the Ward identity at the presence of a boundary (see Eq.\ (24) of \cite{CARDY2006333})
    \begin{align}
    \begin{split} \label{eq:Ward}
        & \Braket{\mathcal{T}(x) \tilde{\mathcal{O}}_{\beta}(z_1) \tilde{\mathcal{O}}_{-\beta}(z_2) }_{\mathbb{H}} \\
        &= \sum_{j=1}^2 \left( \frac{\Delta(\beta)}{(x-z_j)^2} + \frac{1}{x-z_j} \partial_{z_j} + \frac{\Delta(\beta)}{(x-\bar{z}_j)^2} + \frac{1}{x-\bar{z}_j} \partial_{\bar{z}_j} \right)
    \Braket{\tilde{\mathcal{O}}_{\beta}(z_1) \tilde{\mathcal{O}}_{-\beta}(z_2)}_{\mathbb{H}},
    \end{split}
    \end{align}
    where $\mathcal{T}$ denotes the boundary stress-energy tensor with $x \in \mathbb{R}$,
    \item the expression (see Eq.\ (2.2) of \cite{Camia_2020})
    \begin{align}
    \begin{split} \label{eq:OO}
    & \Braket{\tilde{\mathcal{O}}_{\beta}(z_1) \tilde{\mathcal{O}}_{-\beta}(z_2)}_{\mathbb{H}} = |z_1-z_2|^{-4\Delta(\beta)} |z_1-\bar{z}_2|^{4\Delta(\beta)} |z_1-\bar{z}_1|^{-2\Delta(\beta)} |z_2-\bar{z}_2|^{-2\Delta(\beta)} \\
    & \qquad \cdot \exp\left[ -2\Delta(\beta)(1-\sigma(z_1,z_2)) {}_3 F_2\big(1, 1, 4/3; 2, 5/3; 1-\sigma(z_1,z_2) \big)\right],
    \end{split}
    \end{align}
    where ${}_3 F_2$ is the hypergeometric function.\footnote{We note that the expression \eqref{eq:OO}, which was derived in \cite{Camia_2020} using results from \cite{han2017brownian,beliaev2013some}, is rigorous.}
\end{itemize}

Plugging \eqref{eq:OO} into the right hand side of \eqref{eq:Ward}, sending $x \to 0$, and comparing with \eqref{eq:OOB} shows that the result equals $\frac{\sqrt{\lambda}}{\hat{c}_b}\Braket{ \mathcal{B}(0) \tilde{\mathcal{O}}_{\beta}(z_1) \tilde{\mathcal{O}}_{-\beta}(z_2) }_{\mathbb{H}}$. By translation invariance, this means that, on the real line, we can make the identification $\mathcal{T} = \frac{\sqrt{\lambda}}{\hat{c}_b}\mathcal{B}$.
According to \eqref{eq:canonical_boundary}, for any two points $x_1,x_2 \in \mathbb{R}$, $\mathcal{T}$ would then have the two-point function
\begin{align}
    \Braket{{\mathcal T}(x_1){\mathcal T}(x_2)}_{\mathbb H} = \frac{\lambda}{\hat{c}_b^2 \, |x_1-x_2|^{4}},
\end{align}
but since the stress-energy tensor $\mathcal{T}$ satisfying \eqref{eq:Ward} is assumed to be normalized in such a way that
\begin{align}
    \Braket{{\mathcal T}(z_1){\mathcal T}(z_2)}_{\mathbb H} = \frac{c/2}{|z_1-z_2|^{4}},
\end{align}
where $c$ denotes the central charge (see, e.g., Eq.\ (8) of \cite{CARDY2006333}), from the relation $c=2\lambda$ valid for the BLS with intensity $\lambda$, we conclude that $\hat{c}_b=1$ and determine the boundary stress-energy tensor to be
\begin{align}
    \mathcal{T} = \sqrt{\lambda}\mathcal{B}.
\end{align}

In other words, the discussion above leads to the following result.
\begin{proposition} \label{th:Ward}
For any choice of distinct points $z_1, z_2 \in \mathbb{H}$ and $x \in \mathbb{R}$, $\mathcal{T}(x) = \sqrt{\lambda}\mathcal{B}(x)$ satisfies the Ward identity \eqref{eq:Ward}.
\end{proposition}

\subsection{Operator product expansion} \label{sec:boundary-stress-energy-tensor-OPE}

In this section we discuss the OPE of $\mathcal{T} \times \mathcal{T}$. In order to do that, we first express the mixed four-point function $\Braket{ \mathcal{B}(x_1) \mathcal{B}(x_2) \tilde{\mathcal{O}}_{\beta}(z_1) \tilde{\mathcal{O}}_{-\beta}(z_2) }_{\mathbb{H}}$ in terms of $\mu^{\text{loop}}_{\mathbb{H}}$-weights. For that purpose, it is useful to introduce some additional notation. We define
\begin{align}
\begin{split}
    \beta^x_z := \lim_{\varepsilon \to 0} \varepsilon^{-2} \mu^{\text{loop}}_{\mathbb{H}}\big( \ell \cap B_{\varepsilon}(x) \neq \emptyset, z \in \bar\ell \big)
\end{split}
\end{align}
and
\begin{align} \label{def:beta3412}
    \beta^{x_1,x_2}_{z_1|z_2} = \beta^{x_1,x_2}_{z_2|z_1} := \lim_{\varepsilon \to 0} \varepsilon^{-4} \mu_{\mathbb{H}}^{\text{loop}}(\ell \cap B_{\varepsilon}(x_j) \neq \emptyset \text{ for } j=1,2; \ell \text{ separates } z_1,z_2),
\end{align}
where the existence of both limits follows from the proof of Lemma 2.2 of \cite{camia2021scalar}.

\begin{proposition} \label{prop:OOBB}
For any choice of points $z_1,z_2 \in \mathbb{H}$ and $x_1,x_2 \in \mathbb{R}$, with the notation of Proposition \ref{prop:OOB}, we have that
\begin{align}
\begin{split} \label{eq:OOBB}
& \Braket{ \mathcal{B}(x_1) \mathcal{B}(x_2) \tilde{\mathcal{O}}_{\beta}(z_1) \tilde{\mathcal{O}}_{-\beta}(z_2) }_{\mathbb{H}} \\
&= \Braket{\tilde{\mathcal{O}}_{\beta}(z_1) \tilde{\mathcal{O}}_{-\beta}(z_2)}_{\mathbb{H}} \left[ \beta^{x_1,x_2} - (1-\cos\beta) \beta^{x_1,x_2}_{z_1|z_2} + \lambda (1-\cos\beta)^2 \beta^{x_1}_{z_1|z_2} \beta^{x_2}_{z_1|z_2} \right] \\
&= \Braket{\tilde{\mathcal{O}}_{\beta}(z_1) \tilde{\mathcal{O}}_{-\beta}(z_2)}_{\mathbb{H}} \left[ \Braket{\mathcal{B}(x_1) \mathcal{B}(x_2)}_{\mathbb{H}} - (1-\cos\beta) \beta^{x_1,x_2}_{z_1|z_2} + \lambda (1-\cos\beta)^2 \beta^{x_1}_{z_1|z_2} \beta^{x_2}_{z_1|z_2} \right].
\end{split}
\end{align}
\end{proposition}

\noindent{\bf Proof.} The result follows from carrying out the analysis leading to (5.6) in \cite{camia2021scalar} to the upper half-plane, with the exponents $2/3$ replaced by $2$. The last line of \eqref{eq:OOBB} follows from Proposition \ref{prop:n-point-functions}, which allows us to make the identification $\Braket{\mathcal{B}(x_1) \mathcal{B}(x_2)}_{\mathbb{H}} = \beta^{x_1,x_2}$. \qed

\medskip

We now discuss \eqref{eq:OOBB} in some detail. First of all, we note that, arguing as in the proof of Proposition \ref{prop:OOB} and using again Proposition 1 of \cite{beliaev2013some}, one can find an explicit expression for the last term between square brackets of the last line of \eqref{eq:OOBB}, namely using
\begin{align}
\begin{split} \label{eq:betaxzz}
    \beta^x_{z_1|z_2} &= \lim_{\varepsilon \to 0} \varepsilon^{-2} \mu^{\text{loop}}_{\mathbb{H}}\big( \ell \cap B_{\varepsilon}(x) \neq \emptyset, \ell \text{ separates } z_1,z_2 \big) = \mu^{\text{bub}}_{\mathbb{H},x}\big(\ell \text{ separates } z_1,z_2 \big) \\
    &= \frac{8}{5} \left( \mu^{\text{SLE}_{8/3}}_{\mathbb{H},x}(z_1 \in \bar\ell) + \mu^{\text{SLE}_{8/3}}_{\mathbb{H},x}(z_2 \in \bar\ell) - 2 \mu^{\text{SLE}_{8/3}}_{\mathbb{H},x}(z_1,z_2 \in \bar\ell) \right) \\
    &= \frac{2}{5} \frac{(\Im z_1)^2}{\left(\text{dist}(x,z_1)\right)^4} + \frac{2}{5} \frac{(\Im z_2)^2}{\left(\text{dist}(x,z_2)\right)^4} - \frac{4}{5} \frac{\Im z_1}{\left(\text{dist}(x,z_1)\right)^2} \frac{\Im z_2}{\left(\text{dist}(x,z_2)\right)^2} G(\sigma(z_1,z_2)),
\end{split}
\end{align}
where $\mu^{\text{bub}}_{\mathbb{H},x}$ is the measure of Brownian bubbles in $\mathbb{H}$ pinned at $x$ and $\mu^{\text{SLE}_{8/3}}_{\mathbb{H},x}$ is the measure of SLE$_{8/3}$ excursions in $\mathbb{H}$ from $x$ (in other words, they are the measures $\mu^{\text{SLE}_{\text{bub}}}_{\mathbb{H}}$ and $\mu^{\text{SLE}_{8/3}}_{\mathbb{H}}$, respectively, after a translation of $\overline{\mathbb{H}}$ that moves the origin to $x$).

Next, we note that the first two terms between square brackets in the last line of \eqref{eq:OOBB} diverge as $|x_1-x_2| \to 0$. This is clear for $\Braket{{\mathcal B}(x_1){\mathcal B}(x_2)}_{\mathbb H} = |x_1-x_2|^{-4}$, so we analyze the remaining term. We need to understand the behavior of $\beta^{x_1,x_2}_{z_1|z_2}$ as $|x_1-x_2| \to 0$. For any $s>0$, the scale invariance of the BLS implies that
\begin{align}
\begin{split}
    \beta^{sx_1,sx_2}_{z_1|z_2} &= \lim_{\varepsilon \to 0} \varepsilon^{-4} \mu_{\mathbb{H}}^{\text{loop}}(\ell \cap B_{\varepsilon}(s x_j) \neq \emptyset \text{ for } j=1,2; \ell \text{ separates } z_1,z_2) \\
    &= s^{-4} \lim_{\varepsilon \to 0} (\varepsilon/s)^{-4} \mu_{\mathbb{H}}^{\text{loop}}(\ell \cap B_{\varepsilon/s}(x_j) \neq \emptyset \text{ for } j=1,2; \ell \text{ separates } z_1/s,z_2/s) \\
    &= s^{-4} \beta^{x_1,x_2}_{\frac{z_1}{s}|\frac{z_2}{s}}.
\end{split}
\end{align}
In particular, $\beta^{0,s}_{z_1|z_2} = s^{-4} \beta^{0,1}_{\frac{z_1}{s}|\frac{z_2}{s}}$.

For $s$ very small, $z_1/s$ and $z_2/s$ are very far from $0$ and $1$, and one can expect that
$\beta^{0,1}_{\frac{z_1}{s}|\frac{z_2}{s}} \sim \beta^0_{\frac{z_1}{s}|\frac{z_2}{s}}$. This can be justified using the relation between $\mu^{\text{loop}}$ and critical percolation (see the proof of Lemma 2.2 in the appendix of \cite{camia2021scalar}). Essentially, $\mu_{\mathbb{H}}^{\text{loop}}$ of the set of loops getting $\varepsilon$-close to 0 and 1 and separating $z_1/s$ from $z_2/s$ can be expressed in terms of the probability that a critical percolation cluster contained in the upper half-plane gets close to $0$ and $1$ and ``swallows'' $z_1/s$ but not $z_2/s$ or vice versa. Because of properties of percolation (namely the fact that a percolation configuration can be obtained by a local exploration process and the fact that two such processes are independent when they happen in spatially separated regions), the probability of such an event in a percolation model with a small mesh size $a$ is well approximated by the product of four terms: (i) the probability that the semi-annulus of inner radius $\varepsilon$ and outer radius $1/2$ centered at 0 contains a three-arm event (i.e., contains a percolation cluster crossing the semi-annulus without touching the real line), (ii) the probability that the semi-disk of inner radius $\varepsilon$ and outer radius $1/2$ centered at 1 contains a three-arm event, (iii) the probability that the clusters crossing the semi-annuli centered at 0 and 1 are connected, (iv) the probability that the cluster crossing the semi-annulus centered at $0$ ``swallows'' $z_1/s$ but not $z_2/s$ or vice versa.
In the scaling limit, $a \to 0$, the first two terms are equal and scale like $\varepsilon^2$: (i)$=$(ii) $\sim \varepsilon^2$. Term (iii) is bounded away from $0$ and $1$ as $a \to 0$ by the well-known (and celebrated) Russo-Seymour-Welsh theorem, so it can be neglected. Since $\beta^{0,1}_{\frac{z_1}{s}|\frac{z_2}{s}}$ is obtained by multiplying $\mu^{\text{loop}}_{\mathbb{H}}$ by $\varepsilon^{-4}$, we can use one factor of $\varepsilon^{-2}$ to compensate for (ii). If we now combine (i) and (iv) and multiply them by the remaining factor of $\varepsilon^{-2}$, as $\varepsilon \to 0$, we obtain $\beta^0_{\frac{z_1}{s}|\frac{z_2}{s}}$. Putting everything together and using the last line of \eqref{eq:betaxzz}, we obtain
$\beta^{0,1}_{\frac{z_1}{s}|\frac{z_2}{s}} \sim \beta^0_{\frac{z_1}{s}|\frac{z_2}{s}} \sim s^2$, which suggests that $\beta^{0,s}_{z_1|z_2} \sim s^{-2}$ as $s \to 0$, and more generally, 
\begin{align} \label{eq:asymptotic-behavior}
    \beta^{x_1,x_2}_{z_1|z_2} \sim |x_1-x_2|^{-2} \text{ as } |x_1-x_2| \to 0.
\end{align}

Combined with \eqref{eq:OOBB}, these observations lead to the conclusion that,
\begin{align}
\begin{split}
\Braket{ \mathcal{B}(x_1) \mathcal{B}(x_2) \tilde{\mathcal{O}}_{\beta}(z_1) \tilde{\mathcal{O}}_{-\beta}(z_2) }_{\mathbb{H}} = \Braket{\tilde{\mathcal{O}}_{\beta}(z_1) \tilde{\mathcal{O}}_{-\beta}(z_2)}_{\mathbb{H}} |x_1-x_2|^{-4} \left[ 1 + O\left( |x_1-x_2|^{2} \right) \right],
\end{split}
\end{align}
which implies that the OPE of $\mathcal{B} \times \mathcal{B}$ starts with
\begin{align} \label{eq:BxB-first-term}
\begin{split}
\mathcal{B}(x) \times \mathcal{B}(x') = \frac{\mathbb{1}}{|x-x'|^{4}}  + \ldots,
\end{split}
\end{align}
where $\mathbb{1}$ denotes the identity operator.

The identification $\mathcal{T} = \sqrt{\lambda} \, \mathcal{B}$, made in the previous section, gives the OPE
\begin{align} \label{eq:TxT-first-term}
\begin{split}
\mathcal{T}(x) \times \mathcal{T}(x') = \frac{\lambda}{|x-x'|^{4}} \, \mathbb{1} + \ldots,
\end{split}
\end{align}
which is consistent with $\mathcal{T}$ being the stress-energy tensor of a CFT with central charge $c=2\lambda$.

In order to guess the next term in the OPE of $\mathcal{T} \times \mathcal{T}$, note that
\begin{align}
\begin{split}
    \beta^{x_1,x_2}_{z_1|z_2} &= \lim_{\varepsilon \to 0} \varepsilon^{-4} \mu_{\mathbb{H}}^{\text{loop}}(\ell \cap B_{\varepsilon}(x_j) \neq \emptyset \text{ for } j=1,2; \ell \text{ separates } z_1,z_2) \\
    &= \lim_{\varepsilon \to 0} \varepsilon^{-4} \mu_{\mathbb{H}}^{\text{loop}}(\ell \cap B_{\varepsilon}(x_1) \neq \emptyset; \ell \text{ separates } z_1,z_2) \\
    &\quad + \lim_{\varepsilon \to 0} \varepsilon^{-4} \Big[ \mu_{\mathbb{H}}^{\text{loop}}(\ell \cap B_{\varepsilon}(x_j) \neq \emptyset \text{ for } j=1,2; \ell \text{ separates } z_1,z_2) \\
    & \qquad - \mu_{\mathbb{H}}^{\text{loop}}(\ell \cap B_{\varepsilon}(x_1) \neq \emptyset; \ell \text{ separates } z_1,z_2) \Big] \\
    &= \lim_{\varepsilon \to 0} \varepsilon^{-4} \Big[ \mu_{\mathbb{H}}^{\text{loop}}(\ell \cap B_{\varepsilon}(x_1) \neq \emptyset; \ell \text{ separates } z_1,z_2) \\
    & \qquad - \mu_{\mathbb{H}}^{\text{loop}}(\ell \cap B_{\varepsilon}(x_1) \neq \emptyset, \ell \cap B_{\varepsilon}(x_2) = \emptyset ; \ell \text{ separates } z_1,z_2) \Big] \\
    &= \lim_{\varepsilon \to 0} \varepsilon^{-2} \mu_{\mathbb{H}}^{\text{loop}}(\ell \cap B_{\varepsilon}(x_1) \neq \emptyset; \ell \text{ separates } z_1,z_2) \\
    & \qquad \cdot \varepsilon^{-2} \Big[ 1 - \frac{\mu_{\mathbb{H}}^{\text{loop}}(\ell \cap B_{\varepsilon}(x_1) \neq \emptyset, \ell \cap B_{\varepsilon}(x_2) = \emptyset ; \ell \text{ separates } z_1,z_2)}{\mu_{\mathbb{H}}^{\text{loop}}(\ell \cap B_{\varepsilon}(x_1) \neq \emptyset; \ell \text{ separates } z_1,z_2)} \Big].
\end{split}
\end{align}
Observing that
\begin{align}
    \lim_{\varepsilon \to 0} \varepsilon^{-2} \mu_{\mathbb{H}}^{\text{loop}}(\ell \cap B_{\varepsilon}(x_1) \neq \emptyset; \ell \text{ separates } z_1,z_2) = \mu_{\mathbb{H},x_1}^{\text{bub}}(\ell \text{ separates } z_1,z_2)
\end{align}
is finite, \eqref{eq:asymptotic-behavior} suggests that, as $\varepsilon \to 0$,
\begin{align}
\begin{split}
    \varepsilon^{-2} \left[ 1 - \frac{\mu_{\mathbb{H}}^{\text{loop}}(\ell \cap B_{\varepsilon}(x_1) \neq \emptyset, \ell \cap B_{\varepsilon}(x_2) = \emptyset ; \ell \text{ separates } z_1,z_2)}{\mu_{\mathbb{H}}^{\text{loop}}(\ell \cap B_{\varepsilon}(x_1) \neq \emptyset; \ell \text{ separates } z_1,z_2)} \right]
\end{split}
\end{align}
converges to a function $f(x_1,x_2)$ such that $\lim_{|x_1-x_2| \to 0} |x_1-x_2|^2 f(x_1,x_2) = \text{const}$. If we make this assumption, we are led to the conclusion that
\begin{align}
\begin{split}
    & \Braket{ \mathcal{B}(x_1) \mathcal{B}(x_2) \tilde{\mathcal{O}}_{\beta}(z_1) \tilde{\mathcal{O}}_{-\beta}(z_2) }_{\mathbb{H}} = \frac{\Braket{\tilde{\mathcal{O}}_{\beta}(z_1) \tilde{\mathcal{O}}_{-\beta}(z_2)}_{\mathbb{H}}}{|x_1-x_2|^{4}} \\
    & \quad + \frac{\text{const}}{\sqrt{\lambda}} \frac{\Braket{ \mathcal{B}(x_1) \tilde{\mathcal{O}}_{\beta}(z_1) \tilde{\mathcal{O}}_{-\beta}(z_2) }_{\mathbb{H}}}{|x_1-x_2|^2} + o(|x_1-x_2|^{-2}),
\end{split}
\end{align}
which implies the OPEs
\begin{align}
\begin{split}
    \mathcal{B}(x) \times \mathcal{B}(x') = \frac{1}{|x-x'|^{4}} \, \mathbb{1} + \frac{\text{const}}{\sqrt{\lambda}} \frac{1}{|x-x'|^2} \, \mathcal{B}(x) + \ldots
\end{split}
\end{align}
and
\begin{align}
\begin{split} \label{TT}
    \mathcal{T}(x) \times \mathcal{T}(x') = \frac{\lambda}{|x-x'|^{4}} \, \mathbb{1} + \frac{\text{const}}{|x-x'|^2} \, \mathcal{T}(x) + \ldots .
\end{split}
\end{align}
The general form of the OPE of the stress energy tensor with itself (see, e.g., Eq.\ (8) of \cite{CARDY2006333}) is consistent with \eqref{TT} and the fact that the central charge of the theory is $2\lambda$, and implies that $\text{const}=2$.

\section{Domain perturbations and boundary condition changing operators} \label{sec:domain-perturbations_bc-ops}

In the first part of this section (Section \ref{sec:domain-perturbations}) we discuss the relation between local boundary deformations and the operator $\mathcal{T}$. In the second part (Section \ref{sec:bcc-ops}), we introduce boundary operators that generate Brownian excursions and show that they behave like \emph{boundary condition changing operators} and how they are related to $\mathcal{T}$.

\subsection{Domain perturbations} \label{sec:domain-perturbations}

Roughly speaking, inserting the stress-energy tensor at $z$ corresponds to applying a conformal transformation that preserves $\infty$ and has a simple pole at $z$. Let $\mathbb{H}_{\varepsilon}$ denote the upper half-plane from which a disk of radius $\varepsilon$ center at the origin has been removed, i.e.\ $\mathbb{H}_{\varepsilon} := \mathbb{H} \setminus B_{\varepsilon}(0)$. The map
\begin{align} \label{eq:conformal-map}
    f_{\varepsilon}(z) = z + \frac{\varepsilon^2}{z}
\end{align}
preserves $\infty$, has a simple pole at $0$, and transforms $\mathbb{H}_{\varepsilon}$ to $\mathbb{H}$ confrormally. (Consequently, it maps $\mathbb{C} \setminus B_{\varepsilon}(0)$ conformally to $\mathbb{C}$.) Therefore, in the limit $\varepsilon \to 0$, mapping $\mathbb{H}_{\varepsilon}$ to $\mathbb{H}$ confrormally, which is equivalent to applying $f_{\varepsilon}$ to $\mathbb{H}_{\varepsilon}$, should in some sense be also equivalent to inserting the boundary stress energy tensor at $0$ in the upper half-plane.

Following ideas from \cite{camia_2016_book} (see Proposition 2.6) and \cite{10.1214/18-EJP200} (see Proposition 4.3), we will show that $\mathcal{T}$ can be used to express the effect of domain perturbations on the two-point function of layering vertex operators, corroborating its identification with the boundary stress-energy tensor. For concreteness, we will work in the upper half-plane.

It follows from the definition \eqref{def:tildeO} of $\tilde{\mathcal{O}}_{\beta}$ and Eq.\ (4.3) of \cite{Camia_2016} that
\begin{align}
\begin{split}
    \frac{\Braket{\tilde{\mathcal{O}}_{\beta}(z_1) \tilde{\mathcal{O}}_{-\beta}(z_2)}_{\mathbb{H}_{\varepsilon}}}{\Braket{\tilde{\mathcal{O}}_{\beta}(z_1) \tilde{\mathcal{O}}_{-\beta}(z_2)}_{\mathbb{H}}} = \exp{\left(\lambda(1-\cos\beta) \mu^{\text{loop}}_{\mathbb{H}}(\ell \cap B_{\varepsilon}(0) \neq \emptyset, \ell \text{ separates } z_1,z_2) \right)},
\end{split}
\end{align}
which we can also write as
\begin{align}
\begin{split}
    & \Braket{\tilde{\mathcal{O}}_{\beta}(z_1) \tilde{\mathcal{O}}_{-\beta}(z_2)}_{\mathbb{H}_{\varepsilon}} - \Braket{\tilde{\mathcal{O}}_{\beta}(z_1) \tilde{\mathcal{O}}_{-\beta}(z_2)}_{\mathbb{H}} \\
    &  = \lambda(1-\cos\beta) \mu^{\text{loop}}_{\mathbb{H}}(\ell \cap B_{\varepsilon}(0) \neq \emptyset, \ell \text{ separates } z_1,z_2) \Braket{\tilde{\mathcal{O}}_{\beta}(z_1) \tilde{\mathcal{O}}_{-\beta}(z_2)}_{\mathbb{H}} \\
    & \qquad + O\left( \left(\mu^{\text{loop}}_{\mathbb{H}}(\ell \cap B_{\varepsilon}(0) \neq \emptyset, \ell \text{ separates } z_1,z_2)\right)^2 \right).
\end{split}
\end{align}

This allows us to define a ``derivative'' with respect to domain perturbations:
\begin{align}
\begin{split}
    & \partial_{D} \Braket{\tilde{\mathcal{O}}_{\beta}(z_1) \tilde{\mathcal{O}}_{-\beta}(z_2)}_{D} \Big|_{D=\mathbb{H}} \\
    &:= \lim_{\varepsilon \to 0} \frac{\Braket{\tilde{\mathcal{O}}_{\beta}(z_1) \tilde{\mathcal{O}}_{-\beta}(z_2)}_{\mathbb{H}} - \Braket{\tilde{\mathcal{O}}_{\beta}(z_1) \tilde{\mathcal{O}}_{-\beta}(z_2)}_{\mathbb{H}_{\varepsilon}}}{\varepsilon^2} \\
    &= \lambda(1-\cos\beta) \Braket{\tilde{\mathcal{O}}_{\beta}(z_1) \tilde{\mathcal{O}}_{-\beta}(z_2)}_{\mathbb{H}_{\varepsilon}} \lim_{\varepsilon \to 0} \varepsilon^{-2} \mu^{\text{loop}}_{\mathbb{H}}(\ell \cap B_{\varepsilon}(0) \neq \emptyset, \ell \text{ separates } z_1,z_2) \\
    &= - \Braket{\mathcal{T}(0) \tilde{\mathcal{O}}_{\beta}(z_1) \tilde{\mathcal{O}}_{-\beta}(z_2)}_{\mathbb{H}},
\end{split}
\end{align}
where the last equality follows from \eqref{eq:OOB-proof} and shows the appearance of the operator $\mathcal{T}$ inserted at 0.
From this expression, using \eqref{eq:OOB} one can obtain an explicit formula for the derivative above.

\medskip

\subsection{Boundary condition changing operators} \label{sec:bcc-ops}

Working for concreteness in the upper half-plane $\mathbb{H}$, in this section we consider a Brownian loop soup in $\mathbb{H}$ with an additional Brownian excursion $\gamma_{0,x}$ in $\mathbb{H}$ from the origin $0$ to a point $x$ on the real line, independent of the loop soup. We think of $\gamma_{0,x}$ as being generated by the insertion of a pair of 
boundary changing operators at $0$ and $x$, and we think of the concatenation of $\gamma_{0,x}$ with the interval $[0,x]$ as forming a loop added to the loop soup, with outer boundary $\text{b}$. The boundary curve $\text{b}$ is assigned an independent random sign $\sigma_{\text{b}}=\pm 1$ with equal probability. In this model, expectations will be denoted by $\Braket{\Phi(0) \vert \cdot \vert \Phi(x)}_{\mathbb{H}}$, where $\Phi(y)$ denotes a boundary operator inserted at $y$. We will also use $\mathbb{P}^{\text{B}}_{\mathbb{H};0,x}$ to denote the probability distribution of a Brownian excursion\footnote{A Brownian excursion in $\mathbb{H}$ from $0$ to $x$ is a Brownian path starting at $0$ and ending at $x$ conditioned to stay in $\mathbb{H}$. Defining such an object requires some care because it implies conditioning on en event of zero probability. However, this can be done and $\mathbb{P}^{\text{B}}_{\mathbb{H};0,x}$ is a well defined probability measure.} in $\mathbb{H}$ from $0$ to $x$ and $\mathbb{E}^{\text{B}}_{\mathbb{H};0,x}$ to denote expectation with respect to $\mathbb{P}^{\text{B}}_{\mathbb{H};0,x}$, while $\mathbf{I}(\cdot)$ will denote the indicator function. With this notation, using \eqref{def:tildeO}, we have
\begin{align}
\begin{split} \label{eq:OO_bc-ops}
    & \Braket{\Phi(0) \left\vert \tilde{\mathcal{O}}_{\beta}(z_1) \tilde{\mathcal{O}}_{-\beta}(z_2) \right\vert \Phi(x)}_{\mathbb{H}} = \lim_{\delta \to 0} (\hat{c}_{\mathbb{H}}\delta)^{-4\Delta(\beta)} \Braket{\exp{\Big[i \beta \Big( \sum_{\stackrel{\ell \in \mathcal{L}^{\delta} }{z_1 \in \bar{\ell}}} \sigma_{\ell} - \sum_{\stackrel{\ell \in \mathcal{L}^{\delta} }{z_2 \in \bar{\ell}}} \sigma_{\ell} \Big) \Big]}}_{\mathbb{H}} \\
    & \qquad \qquad \qquad \qquad \qquad \qquad \qquad \cdot \sum_{\sigma_{\text{b}}=\pm 1} \frac{1}{2} \mathbb{E}^{\text{bub}}_{\mathbb{H};0,x} \left[ e^{i\beta\sigma_{\text{b}}\mathbf{I}(z_1 \in \bar{\text{b}})} e^{-i\beta\sigma_{\text{b}}\mathbf{I}(z_2 \in \bar{\text{b}})} \right] \\
    & = \Braket{\tilde{\mathcal{O}}_{\beta}(z_1) \tilde{\mathcal{O}}_{\beta}(z_2)}_{\mathbb{H}} \left( \mathbb{P}^{\text{B}}_{\mathbb{H};0,x}(\text{b does not separate } z_1,z_2)
    + \cos\beta \mathbb{P}^{\text{B}}_{\mathbb{H};0,x}(\text{b separates } z_1,z_2) \right) \\
    & = \Braket{\tilde{\mathcal{O}}_{\beta}(z_1) \tilde{\mathcal{O}}_{\beta}(z_2)}_{\mathbb{H}} \left[ 1 - (1-\cos\beta) \mathbb{P}^{\text{B}}_{\mathbb{H};0,x}(\text{b separates } z_1,z_2) \right].
\end{split}
\end{align}

Now let $\Gamma_{0,x}$ denote an SLE$_{8/3}$ in the upper half-plane from $0$ to $x$, with distribution $\mathbb{P}^{\text{SLE}_{8/3}}_{\mathbb{H};0,x}$. The equivalence $\mu^{\text{bub}}_{\mathbb{H}} = \frac{8}{5} \mu^{\text{SLE}_{8/3}}_{\mathbb{H}}$ and the relation between $\mathbb{P}^{\text{B}}_{\mathbb{H};0,x}$ and $\mu^{\text{bub}}_{\mathbb{H}}$ on the one hand and between $\mathbb{P}^{\text{SLE}_{8/3}}_{\mathbb{H};0,x}$ and $\frac{8}{5} \mu^{\text{SLE}_{8/3}}_{\mathbb{H}}$, on the other, imply that
\begin{align} \label{eq:equivalence-probabilities}
    \mathbb{P}^{\text{B}}_{\mathbb{H};0,x}(\text{b separates } z_1,z_2) = \mathbb{P}^{\text{SLE}_{8/3}}_{\mathbb{H};0,x}(\Gamma_{0,x} \text{ separates } z_1,z_2),
\end{align}
which gives
\begin{align}
\begin{split} \label{eq:OO_bc-ops_bis}
    & \Braket{\Phi(0) \left\vert \tilde{\mathcal{O}}_{\beta}(z_1) \tilde{\mathcal{O}}_{-\beta}(z_2) \right\vert \Phi(x)}_{\mathbb{H}} \\
    &= \Braket{\tilde{\mathcal{O}}_{\beta}(z_1) \tilde{\mathcal{O}}_{-\beta}(z_2)}_{\mathbb{H}} \left[ 1 - (1-\cos\beta) \mathbb{P}^{\text{SLE}_{8/3}}_{\mathbb{H};0,x}(\Gamma_{0,x} \text{ separates } z_1,z_2) \right].
\end{split}
\end{align}

This leads to two interesting conclusions. On the one hand, using (3.2) and (3.3) from Proposition 1 of \cite{beliaev2013some}, and letting $\bar\Gamma_{0,x}$ denote the interior of the loop formed by the concatenation of $\Gamma_{0,x}$ with $[0,x]$, we have
\begin{align}
\begin{split}
    & \mathbb{P}^{\text{SLE}_{8/3}}_{\mathbb{H};0,x}(\Gamma_{0,x} \text{ separates } z_1,z_2) \\
    & = \mathbb{P}^{\text{SLE}_{8/3}}_{\mathbb{H};0,x}(z_1 \in {\bar\Gamma_{0,x}}) + \mathbb{P}^{\text{SLE}_{8/3}}_{\mathbb{H};0,x}(z_2 \in {\bar\Gamma_{0,x}}) -2 \mathbb{P}^{\text{SLE}_{8/3}}_{\mathbb{H};0,x}(z_1,z_2 \in {\bar\Gamma_{0,x}}) \\
    & = \frac{5}{8} \left\{ \frac{2}{5} \left[ \left(\Im\left(\frac{1}{z_1}\right)\right)^2 + \left(\Im\left(\frac{1}{z_2}\right)\right)^2 \right] - \frac{4}{5} \Im\left(\frac{1}{z_1}\right) \Im\left(\frac{1}{z_2}\right) G(\sigma(z_1,z_2)) \right\} x^2 + o(x^2),
\end{split}
\end{align}
where $G$ and $\sigma$ are the same as in Proposition \ref{prop:OOB}.
Plugging this expression into \eqref{eq:OO_bc-ops_bis} and using Proposition \ref{prop:OOB} (remembering that $\hat{c}_b=1$ and $\mathcal{T}=\sqrt{\lambda}\mathcal{B}$) gives
\begin{align}
\begin{split} \label{eq:PhiOOPhi}
    & \Braket{\Phi(0) \left\vert \tilde{\mathcal{O}}_{\beta}(z_1) \tilde{\mathcal{O}}_{-\beta}(z_2) \right\vert \Phi(x)}_{\mathbb{H}} \\
    &= \Braket{\tilde{\mathcal{O}}_{\beta}(z_1) \tilde{\mathcal{O}}_{-\beta}(z_2)}_{\mathbb{H}} + \frac{5}{8} \Braket{ \mathcal{T}(0) \tilde{\mathcal{O}}_{\beta}(z_1) \tilde{\mathcal{O}}_{-\beta}(z_2) }_{\mathbb{H}} \frac{x^2}{\lambda} + o(x^2),
\end{split}
\end{align}
showing the appearance of the boundary stress-energy tensor.

On the other hand, it follows from the analysis in Section VI of \cite{Simmons_2009} that\footnote{The analysis in Section VI of \cite{Simmons_2009} is carried out in a strip but, as the authors point out at the end of the section, the quantities involved are conformally invariant, so the results apply to the upper half-plane as well.}
\begin{align}
\begin{split} \label{eq:equivalence-prob-correlations}
    1-2\mathbb{P}^{\text{SLE}_{8/3}}_{\mathbb{H};0,x}(\Gamma_{0,x} \text{ separates } z_1,z_2) = \frac{\Braket{\phi_{1,2}(0) \phi_{1,2}(x) \phi_{2,1}(z_1) \phi_{2,1}(z_2)}^{O(0)}_{\mathbb{H}}}{\Braket{\phi_{1,2}(0) \phi_{1,2}(x)}^{O(0)}_{\mathbb{H}}},
\end{split}
\end{align}
where the expectations in the right hand side are computed in the $O(n \to 0)$ model in the upper half-plane and $\phi_{1,2}(x)$ is a \emph{boundary condition changing operator} applied at $x \in \mathbb{R}$ and $\phi_{2,1}(z)$ is a \emph{twist operator} applied at $z \in \mathbb{H}$. Combined with \eqref{eq:OO_bc-ops_bis}, \eqref{eq:equivalence-prob-correlations} gives, for $\beta=\pi$,
\begin{align} \label{eq:OO_bc-ops_tris}
\begin{split}
    & \Braket{\Phi(0) \left\vert \tilde{\mathcal{O}}_{\pi}(z_1) \tilde{\mathcal{O}}_{\pi}(z_2) \right\vert \Phi(x)}_{\mathbb{H}} \\
    &= \Braket{\tilde{\mathcal{O}}_{\pi}(z_1) \tilde{\mathcal{O}}_{\pi}(z_2)}_{\mathbb{H}} \left[ 1 - 2 \mathbb{P}^{\text{SLE}_{8/3}}_{\mathbb{H};0,x}(\tilde{\gamma}_{0,x} \text{ separates } z_1,z_2) \right] \\
    &= \Braket{\tilde{\mathcal{O}}_{\pi}(z_1) \tilde{\mathcal{O}}_{\pi}(z_2)}_{\mathbb{H}} \frac{\Braket{\phi_{1,2}(0) \phi_{1,2}(x) \phi_{2,1}(z_1) \phi_{2,1}(z_2)}^{O(0)}_{\mathbb{H}}}{\Braket{\phi_{1,2}(0) \phi_{1,2}(x)}^{O(0)}_{\mathbb{H}}},
\end{split}
\end{align}
which leads to an interesting equivalence between quantities in the BLS conformal field theory and in the $O(0)$ model, namely
\begin{align} \label{eq:equivalence}
\begin{split}
    & \frac{\Braket{\Phi(0) \left\vert \tilde{\mathcal{O}}_{\pi}(z_1) \tilde{\mathcal{O}}_{\pi}(z_2) \right\vert \Phi(x)}_{\mathbb{H}}}{\Braket{\tilde{\mathcal{O}}_{\pi}(z_1) \tilde{\mathcal{O}}_{\pi}(z_2)}_{\mathbb{H}}} = \frac{\Braket{\phi_{1,2}(0) \phi_{1,2}(x) \phi_{2,1}(z_1) \phi_{2,1}(z_2)}^{O(0)}_{\mathbb{H}}}{\Braket{\phi_{1,2}(0) \phi_{1,2}(x)}^{O(0)}_{\mathbb{H}}}.
\end{split}
\end{align}
Note that the equivalence is valid for all $\lambda\geq0$.

If we now let $\lambda \to 0$, all loops from the loop soup are suppressed and
\begin{align}
\begin{split}
    \lim_{\lambda \to 0} \Braket{\tilde{\mathcal{O}}_{\pi}(z_1) \tilde{\mathcal{O}}_{\pi}(z_2)}_{\mathbb{H}} = 1,
\end{split}
\end{align}
and we get
\begin{align}
\begin{split}
    \Braket{\Phi(0) \left\vert \tilde{\mathcal{O}}_{\pi}(z_1) \tilde{\mathcal{O}}_{\pi}(z_2) \right\vert \Phi(x)}_{\mathbb{H}}^{\lambda=0} = \frac{\Braket{\phi_{1,2}(0) \phi_{1,2}(x) \phi_{2,1}(z_1) \phi_{2,1}(z_2)}^{O(0)}_{\mathbb{H}}}{\Braket{\phi_{1,2}(0) \phi_{1,2}(x)}^{O(0)}_{\mathbb{H}}}.
\end{split}
\end{align}
When $\beta=\pi$, the operators $\tilde{\mathcal{O}}_{\beta}=\tilde{\mathcal{O}}_{\pi}$ act exactly like the twist operators $\phi_{2,1}$. This suggests that we can think of the boundary operator $\Phi$ that inserts a Brownian excursion as a boundary condition changing operator $\phi_{1,2}$ inserting a chordal SLE$_{8/3}$.
More precisely, we make the identification
\begin{align} \label{eq:identification}
\begin{split}
    \Braket{\Phi(0) \left\vert \, \cdot \, \right\vert \Phi(x)}^{\lambda=0}_{\mathbb{H}} = \frac{\Braket{\phi_{1,2}(0) \phi_{1,2}(x) \, \cdot \,}^{\lambda=0}_{\mathbb{H}}}{\Braket{\phi_{1,2}(0) \phi_{1,2}(x)}^{\lambda=0}_{\mathbb{H}}},
\end{split}
\end{align}
where $\phi_{1,2}$ is interpreted as a primary operator with scaling dimension $\Delta=5/8$ (corresponding to the ``1-leg operator'' in the upper half plane with $\Delta(\kappa)=\frac{6-\kappa}{2\kappa}$ for $\kappa=8/3$). Inserting \eqref{eq:identification} into \eqref{eq:PhiOOPhi} leads to
\begin{align}
\begin{split} \label{eq:phiphiOO}
    & \frac{\Braket{\phi_{1,2}(0) \phi_{1,2}(x) \tilde{\mathcal{O}}_{\pi}(z_1) \tilde{\mathcal{O}}_{\pi}(z_2)}^{\lambda=0}_{\mathbb{H}}}{\Braket{\phi_{1,2}(0) \phi_{1,2}(x)}^{\lambda=0}_{\mathbb{H}}} \\
    & \quad = 1 + \frac{5}{8} x^2 \lim_{\lambda \to 0}\frac{1}{\lambda} \Braket{ \mathcal{T}(0) \tilde{\mathcal{O}}_{\pi}(z_1) \tilde{\mathcal{O}}_{\pi}(z_2) }_{\mathbb{H}} + o(x^2),
\end{split}
\end{align}
which corresponds to the OPE
\begin{align}
\begin{split} \label{eq:OPEphiphi}
    \phi_{1,2}(0) \times \phi_{1,2}(x) = \lvert x \rvert^{-5/4} \left( \mathbb{1} + \frac{5}{8} \lim_{\lambda \to 0}\frac{\mathcal{T}(0)}{\lambda} x^2 + o(x^2) \right),
\end{split}
\end{align}
showing how the boundary stress-energy tensor emerges in the limit in which two boundary condition changing operators are brought together.

We note that $\frac{1}{\lambda} \Braket{ \mathcal{T}(0) \tilde{\mathcal{O}}_{\pi}(z_1) \tilde{\mathcal{O}}_{\pi}(z_2) }_{\mathbb{H}}$ is independent of $\lambda$, so that the limit in \eqref{eq:phiphiOO} is not necessary. It is tempting to conjecture that one can remove the limit also from \eqref{eq:OPEphiphi}, obtaining an OPE for $\phi_{1,2} \times \phi_{1,2}$ valid for all values of $\lambda>0$. The resulting expression
\begin{align}
\begin{split}
    \phi_{1,2}(0) \times \phi_{1,2}(x) & = \lvert x \rvert^{-5/4} \left( \mathbb{1} + \frac{5}{8} \frac{\mathcal{T}(0)}{\lambda} x^2 + o(x^2) \right) \\
    & = \lvert x \rvert^{-5/4} \left( \mathbb{1} + \frac{2\Delta}{c} \mathcal{T}(0) x^2 + o(x^2) \right)
\end{split}
\end{align}
with $\Delta=5/8$ and $c=2\lambda$ is consistent with an OPE for primary operators with scaling dimension $5/8$, with structure constant $C^{\mathcal{T}}_{\phi_{1,2},\phi_{1,2}}=1$ for all $\lambda$, see \eqref{OPE}.

It was argued in \cite{Camia_2020} that the $n \to 0$ limit of the $O(n)$ model, which describes single self-avoiding loops, is closely related to the the BLS in the $\lambda \to 0$ limit.
We saw additional evidence for this identification in this section, where we identified the boundary changing operator $\Phi$ with the $O(n)$ model operator $\phi_{1,2}$ and presented evidence that this identification holds for finite $\lambda$.

It is curious to note that a similar identification can be made for $\mathcal{T}$. Its scaling dimension, $\Delta = 2$, coincides with that of a ``2-leg'' boundary operator in the $O(0)$ model, namely $\Delta(\kappa) = (8-\kappa)/\kappa$ for $\kappa = 8/3$.
Geometrically, $\mathcal{T}$ counts the number of loops coming close to its insertion point on the boundary (see Figure \ref{edge_edge}). This geometric interpretation coincides with that of the 2-leg operator, which acts as a sink or source of two self-avoiding random walks at its insertion point that avoid each other and the boundary of the domain. Moreover, although the ensemble of loops of the BLS contains an infinite number of loops, if one considers the set of points $x$ on the real line such that the semi-disk of radius $\varepsilon$ centered at $x$ intersects a ``macroscopic'' loop, its fractal dimension goes to $0$ as $\varepsilon \to 0$, regardless of the value of $\lambda$. This allows us to conjecture that $\mathcal{T}$ can be identified with the 2-leg operator for all $\lambda$. Additionally, it gives an explanation for why the scaling dimension of $\mathcal{B}$ (and $\mathcal{T}$) is independent of $\lambda$.

The discussion in this section can in principle be extended to operators other than $\tilde{\mathcal{O}}_{\beta}$ and to the case in which multiple pairs of boundary condition changing operators are inserted at the boundary.

\section{Conclusions and future perspectives} \label{sec:conclusions}

For conformal field theories associated to the BLS, we have provided integral expressions of the bulk stress-energy tensor $T$ based on the OPE of $\mathcal{E} \times \mathcal{E}$ and of $\mathcal{O}_{\beta} \times \mathcal{O}_{-\beta}$. These expressions can be used to compute certain correlation functions involving $T$ itself, and we have done this for two specific examples, verifying that the corresponding Ward identities are satisfied.  It would be interesting to extend this analysis to $J$, the dimension $(1,0)$ current that also occurs in  these OPEs (see Figure \ref{checkerboard}, position (3,0)).

Furthermore, we have identified a new operator $\mathcal{T}$ whose properties are consistent with those of a boundary stress-energy tensor. The operator $\mathcal{T}$ is essentially a boundary version of the edge counting operator introduced in \cite{camia2021scalar}, and is reminiscent of the boundary stress-energy tensor discussed in \cite{FRIEDRICH2002947,Doyon_2006}. A full verification that $\mathcal{T}$ is the boundary stress-energy tensor, for example by verifying all Ward identities involving $\mathcal{T}$, is an interesting open problem.

We have shown that the insertion of $\mathcal{T}$ is linked to local deformations of the boundary of the domain. This is in the spirit of the discussion at the end of Section 6 of \cite{2005math.....11605W}, where it is argued that the measure on simple loops induced by $\mu^{\text{loop}}$ is well suited to the study of local deformations of the complex structure of Riemann surfaces. It would indeed be interesting to study the CFTs arising from the BLS on Riemann surfaces, in particular in the case of the torus.

The operator $\mathcal{T}$ can also be linked to the insertion of a pair of boundary operators that generate a Brownian excursion between the insertion points. These operators appear to behave like boundary condition changing operators with scaling dimension $\Delta=5/8$, consistent with $\Delta(\kappa)=\frac{6-\kappa}{2\kappa}$ for $\kappa=8/3$. The appearance of the value $\kappa=8/3$ is not surprising since the outer boundary of a Brownian loop is locally distributed like an SLE$_{8/3}$ curve. It would also be interesting to further explore the properties of these putative boundary condition changing operators.

\bigskip

\acknowledgments
The work of M.K.\ is partially supported by the NSF through the grant PHY-1820814.

\bibliographystyle{unsrt}
\bibliography{bibliography}

\end{document}